\documentclass[11pt]{iopart}
\usepackage{graphicx}
\usepackage{mathbbol}
\usepackage{color}
\begin{document}

\title[High-precision phase diagram of spin glasses]{High-precision phase diagram of spin glasses from duality analysis with real-space renormalization and graph polynomials}
\author{Masayuki Ohzeki${}^1$ and Jesper Lykke Jacobsen${}^{2,3}$}
\address{${}^1$Department of Systems Science, Graduate School of Informatics, Kyoto University, 36-1 Yoshida Hon-machi, Sakyo-ku, Kyoto, 606-8501, Japan}
\address{${}^2$LPTENS, \'Ecole Normale Sup\'erieure, 24 rue Lhomond, 75231 Paris, France}
\address{${}^3$Universit\'e Pierre et Marie Curie, 4 place Jussieu, 75252 Paris, France}

\begin{abstract}

We propose a duality analysis for obtaining the critical manifold of
two-dimensional spin glasses. Our method is based on the computation
of quenched free energies with periodic and twisted periodic boundary conditions on a
finite basis. The precision can be systematically improved by increasing the size of the basis,
leading to very fast convergence towards the thermodynamic limit. We apply the method to
obtain the phase diagrams of the random-bond Ising model and $q$-state Potts gauge
glasses. In the Ising case, the Nishimori point is found at $p_N = 0.10929 \pm 0.00002$,
in agreement with and improving on the precision of existing numerical estimations.
Similar precision is found throughout the high-temperature part of the phase diagram.
Finite-size effects are larger in the low-temperature region, but our results are in qualitative
agreement with the known features of the phase diagram. In particular we show analytically 
that the critical point in the ground state is located at finite $p_0$.

\end{abstract}

\ead{mohzeki@i.kyoto-u.ac.jp, jesper.jacobsen@ens.fr}
\submitto{J.\ Phys.\ A: Math.\ Theor.}
\maketitle
\normalsize

\section{Introduction}

Spin glasses exhibit rich phenomena due to the competition between disorder and
frustration effects. Following the pioneering work by Sherrington and Kirkpatrick \cite{Sherrington1975},
the nature of spin glasses is now well understood at the mean-field level \cite{MPV87}.
On the other hand, analytical results on spin glasses in finite dimensions are very scarce,
and much of our understanding relies on numerical investigations. In particular, the
exchange Monte-Carlo method has overcome certain difficulties in computing physical
quantities for finite-dimensional spin glasses
\cite{Binder1986,Swendsen1987,Hukushima1996,Ozeki2007,Ohzeki2010a}.

An important first step in the study of any given statistical model is to determine its phase diagram,
physically characterize the various phases, and precisely locate the critical points.
Only when this has been accomplished can one proceed towards a more refined understanding,
including the determination of critical exponents and the flows between the various critical points.

In the case of spin glasses, the determination of the phase diagram often conceals substantial
difficulties. For instance, the question whether a spin glass phase exists in simple two-dimensional
models, such as the random-bond Ising model, remained controversial for quite some time.
The same can be said about the possibility of having reentrant behaviour at low temperature.
Numerical studies often dedicate a substantial part of the computing time to the precise determination of the phase diagram \cite{Honecker2001,Jacobsen2002}.

In the present study we pave the road towards a better understanding of finite-dimensional
spin glasses by proposing a new analytical method for determining the phase diagram and
critical points of two-dimensional models. We exemplify our method by applying it to a variety of models, including the $\pm J$ random-bond Ising model, the bond-diluted Ising
model, and $q$-state Potts gauge glasses.

The basic idea is the duality transformation, which, in the absence of randomness,
can lead to the exact location of the critical point of the Ising model \cite{Kramers1941}
and its $q$-state generalizations \cite{Wu1976}. There has been a number of recent attempts to generalize the duality approach to inhomogeneous models.
In particular, for the random-bond Ising model the duality transformation combined with the replica method has been successful in deriving an approximate location $(p_N,T_N)$ of the critical point known as the Nishimori point \cite{Nishimori2002,Mailard2003}.%
\footnote{This Nishimori point is situated in a special subspace, the so-called Nishimori line, where the model enjoys a gauge symmetry \cite{Nishimori1981,Nishimori2001}.}
This computation is mean-field like, in the sense that it involves the summation over a single spin in a specified environment.

One of the present authors proposed the conjunctive use of real-space renormalization
to improve the precision with which the replica method estimates the critical points of spin glasses \cite{Ohzeki2008,Ohzeki2009a}. This proposal involves the summation over several spins, corresponding
to a finite part of the inhomogeneous lattice, henceforth referred to as the basis. The size $L$
of the basis determines the degree of the real-space renormalization, and the precision is found
to improve upon increasing $L$.
This method has provided analytical evidence of the absence of the spin glass transition in strong
disorder regions in particular two-dimensional systems \cite{Ohzeki2009b} and derived the critical
points in various systems of spin glasses
\cite{Ohzeki2009c,Nishimoriproc2010,Hector2012, Ohzeki2012a,Ohzeki2012c,Ohzeki2013a,Ohzekiproc2013}.
However, the proposed method still suffers from some shortcomings. 
For instance, for the $\pm J$ Ising model, the results for the slope of the phase boundary near the
pure Ising point do not appear to converge quickly to the known exact answer \cite{Domany1979}
as the degree of the real-space renormalization $L$ is increased \cite{Ohzeki2011a}.
Moreover the estimated phase boundary is not consistent with existing results in
the low-temperature region.

The other author has participated in an independent development of the duality method,
in which the roots of a certain graph polynomial, called the critical polynomial, were
shown to provide good approximations to the critical points of pure spin systems.
A particular motivation for this research was provided by Wu's homogeneity assumption
\cite{Wu1979} which is known to lead to very precise, albeit not exact, approximations for
the critical manifold of the $q$-state Potts model. The Potts critical polynomial $P_B(q,v)$
was initially defined from a contraction-deletion principle \cite{Jacobsen2012}, following earlier
work on the special case ($q=1$) of bond percolation \cite{Scullard2008}. It was subsequently
expressed more simply as a linear combination of partition functions with certain topological
boundary conditions \cite{Scullard2012b,Jacobsen2013}. With hindsight, there are several
analogies with the duality analysis with real-space renormalization. The partition functions are
again defined with respect to a finite part of the lattice, called a basis of size $L$, and the
original homogeneity approach \cite{Wu1979} corresponds to the smallest case $L=1$.
In the context of pure systems, the algorithmic advances in computing $P_B(q,v)$ by a transfer
matrix approach \cite{Jacobsen2013} have brought very large bases within reach, leading to
the determination of critical points with $12$ or $13$ digit accuracy \cite{Jacobsen2014}. On the
other hand, the method has also been shown to produce
good approximations when the coupling constants are inhomogeneous
\cite{Scullard2008,Jacobsen2012}, as is the case in a random system. The graph
polynomial method was however not previously applied to quenched random systems.

It thus appears very natural to compare the two methods carefully, and try to combine the
proven ability of the first method to deal with quenched random system with the second
method's potential for producing very precise results. In the present study we achieve this goal.
Constraining to integer values of $q$, we shall see that the two methods actually differ only
by the boundary conditions applied along the boundary of the basis. In the duality analysis with
real-space renormalization the boundary spins are fixed, whilst the graph polynomial is
a combination of periodic and twisted periodic boundary conditions. The latter choice leads
to far better convergence properties, as will be shown below. Modifying the first method
by imposing periodic rather than fixed boundary conditions, we shall show that it becomes
formally equivalent to the second.

The combined method lifts the shortcomings of the previous real-space renormalization
approach. The phase diagrams and critical points are determined with higher accuracy,
due to the faster convergence in $L$. In particular, we determine the Nishimori point in
the $\pm J$ Ising model and $q=3,4$ Potts gauge glasses with a precision that surpasses
that of state of the art numerical studies.
In addition, we analyze the asymptotic structure of the estimated phase boundary in the
low-temperature
region. In particular, we prove analytically the existence of a non-trivial critical point at
zero temperature, in agreement with various numerical computations.
The compatibility with the existing results of the structure of the phase boundary in both
the high and low-temperature regions serves as a strong validation of our method.

The paper is organized as follows.
The spin glass models to be studied are defined in section \ref{sec2}.
In section \ref{sec3}, we briefly review the previous version of the method, namely the duality analysis
with real-space renormalization. We then modify the boundary conditions in a way that makes
it equivalent to the graph polynomial method. The latter is reviewed in section \ref{sec4}.
The new criterion for criticality is derived and discussed in section~\ref{sec5}, and in section~\ref{sec6}
we then apply it to produce the phase diagrams of 
the various spin glass models under consideration. We test in particular
its ability to precisely locate the Nishimori point and to provide accurate estimates of the
slope of the phase boundary near the pure Ising point.
In the last section we summarize our study and gather a few concluding remarks.

\section{Models}
\label{sec2}

In the following we shall study three different spin glass models. For simplicity, we suppose
each of them defined on the two-dimensional square lattice.

\subsection{Random-bond Ising model}

The simplest and most well-studied model is the $\pm J$ random-bond Ising model
with Hamiltonian
\begin{equation}
H = - \sum_{\langle ij \rangle}J_{ij} S_iS_j \,,
\end{equation}
where $S_i = \pm 1$ is an Ising spin variable located at each site on a lattice, and $J_{ij}$ denotes the interaction between nearest neighbouring pairs represented by $\langle ij \rangle$.
We set $J_{ij} = K \tau_{ij}$, where $K=1/T$ is the inverse temperature and the signs
$\tau_{ij} = \pm 1$ are drawn from the binary distribution
\begin{equation}
P(\tau_{ij}) = (1-p)\delta(\tau_{ij} - 1) + p \delta(\tau_{ij} + 1) \,.
\label{RBIM-dist}
\end{equation}
The couplings are thus frustrated (antiferromagnetic) with probability $p$.

\begin{figure}[tb]
\begin{center}
\includegraphics[width=100mm]{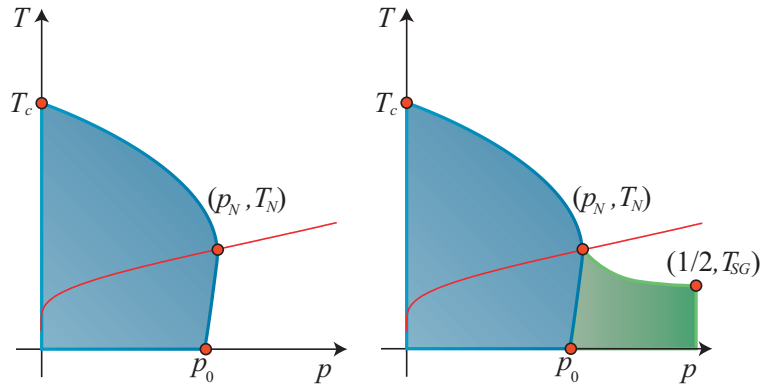}
\end{center}
\caption{Phase diagram of the $\pm J$ random-bond Ising model.
The left panel shows the case of the two-dimensional system, and the right one is
for higher dimensions. The Nishimori line is shown as a red curve.
The shaded (blue) region describes the ferromagnetic phase and the other (white) region
is the paramagnetic phase.
In dimension $d > 2$, one observes a non-trivial low-temperature phase (shown in green), the spin glass phase.
The pure Ising point is represented by $T_c$ and the critical point in the ground state is denoted by $p_0$.
The Nishimori point is situated at the intersection between the Nishimori line and the phase boundary,
at parameters $(p_N,T_N)$. The spin glass transition point is given as $T_{\rm SG}$.
}
\label{paper_fig1}
\end{figure}

The model just introduced has been well investigated both analytically and numerically.
The structure of the phase diagram in the $(p,T)$ plane is well known to be
as in Figure~\ref{paper_fig1}. We now discuss it in some detail,
providing thus a short review of the existing results, in particular those obtained from
the duality analysis with real-space renormalization.

The standard duality analysis leads to the exact value of the critical point of the pure Ising model
($p=0$), namely \cite{Kramers1941}
\begin{equation}
 \exp(-2K_{\rm c} ) = \sqrt{2}-1 \,. \label{Isingselfdual}
\end{equation}
A perturbative calculation to first order in $p$ leads to the exact slope of the phase
boundary at the pure Ising point \cite{Domany1979}:
\begin{equation}
 \frac{1}{T_c}\frac{{\rm d}T}{{\rm d}p} = \frac{\sqrt{2}}{K_{\rm c}} \approx 3.20911 \,.
 \label{slope}
\end{equation}
The main analytical result for $p>0$ is the existence of a special subspace $K=K(p)$, the
so-called Nishimori line
\begin{equation}
 {\rm e}^{-2K(p)} = \frac{p}{1-p} \,, \label{Nishimoriline}
\end{equation}
along which the model possesses a gauge symmetry that makes it possible in particular to
compute exactly the internal energy and establish the pairwise equality of disorder-averaged
moments of the spin-spin correlation function \cite{Nishimori1981,Nishimori2001}.

The duality analysis is not straightforwardly applicable to random spin systems, but the replica
method makes it possible \cite{Nishimori2002,Mailard2003}, up to an unproven assumption
(see section~\ref{sec3}), to determine the location of the critical point.
The concept of real-space renormalization provides a generalization of this computation
using certain bases of size $L$.
Upon increasing $L$, the determination of the high-temperature part of the phase boundary
appears to converge to its true result, still using a similar assumption. This convergence
can be thought to validate the assumption, albeit of course not proving it in a mathematical sense.
In particular, the method provides a precise estimation of the Nishimori point $p_N$
\cite{Ohzeki2009a}. In addition, it shows that a non-trivial low-temperature ordered state, namely the spin glass phase, does not exist in a finite-temperature region in two-dimensional systems \cite{Ohzeki2009b}.

\begin{figure}[t]
\begin{center}
\includegraphics[width=70mm]{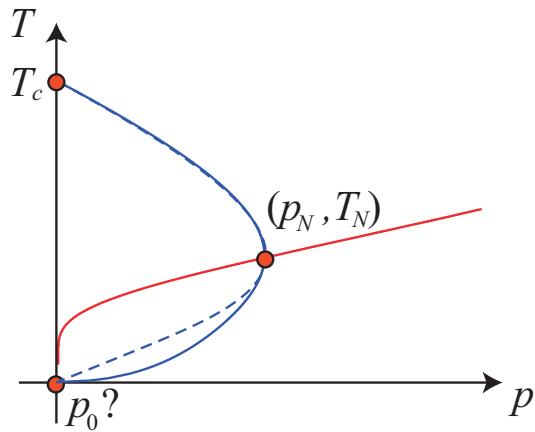}
\end{center}
\caption{Estimated phase diagram by the duality without and with real-space renormalization.
The red curve stands for the Nishimori line.
The solid (blue) curve denotes the result by the duality analysis with real-space renormalization,
corresponding to a basis of size $L=1$.
The dashed curve represents the standard duality analysis using only the replica method.
Neither result captures the non-trivial critical point in the ground state.}
\label{paper_fig2}
\end{figure}

The real-space renormalization, however, has a few shortcomings.
In particular, it predicts the critical points at zero temperature to reside
at a trivial and unacceptable location, $p_0 = 0$,  as shown in Figure~\ref{paper_fig2}.
This is at odds with  several numerical computations that provide non-trivial values of $p_0$
\cite{Amoruso2004,Jinuntuya2012,Fujii2012}, and also with a recent analytical computation
giving $p_0 \approx 0.1031$ \cite{Miyazaki2013}.
In addition, the method does not seem capable of reproducing the value (\ref{slope}) of
the phase boundary slope.

Below we shall present an improvement of the duality analysis with real-space renormalization
that clears these difficulties. In particular, the new method provides
fast convergence to the exact value of the slope in the limit of large $L$.
We shall also show analytically that the new method predicts a finite value of $p_0$, in
agreement with the results just mentioned.

\subsection{Bond-diluted Ising model}

In the bond-diluted Ising model the couplings are either ferromagnetic or absent.
Accordingly, the distribution (\ref{RBIM-dist}) is replaced by
\begin{equation}
 P(\tau_{ij}) = (1-p) \delta(\tau_{ij} - 1) + p \delta(\tau_{ij}) \,.
\end{equation}
The slope of the phase boundary can still be computed exactly from perturbation theory,
and (\ref{slope}) is replaced by \cite{Domany1978}
\begin{equation}
 \frac{1}{T_c}\frac{{\rm d}T}{{\rm d}p} \approx 1.3293 \,.
 \label{dilute-slope}
\end{equation}

\subsection{Potts gauge glass}

The random-bond Ising model can be generalized to a $q$-state model with Hamiltonian
\begin{equation}
 H = - K_{\rm Potts} \sum_{\langle ij \rangle} \delta_q (S_i - S_j + \tau_{ij}) \,,
 \label{H_PGG}
\end{equation}
where $S_i = 1,2,\ldots,q$ and the mod-$q$ Kronecker symbol is defined by
$\delta_q(x) = 1$ if $x=0$ mod $q$, and zero otherwise.
The randomness now takes the form of random twists $\tau_{ij}$ drawn from the
distribution
\begin{equation}
 P(\tau_{ij}) = \big( 1 - (q-1) p \big) \delta(\tau_{ij}) + p \sum_{\tau=1}^{q-1} \delta(\tau_{ij} - \tau) \,,
\end{equation}
with $p \in [0,(q-1)^{-1}]$ controlling the strength of the randomness.

It is easy to see that for $q=2$ this reduces to the random-bond Ising model, with the usual
correspondence $K_{\rm Potts} = 2 K$.

\begin{figure}[tb]
\begin{center}
\includegraphics[width=50mm]{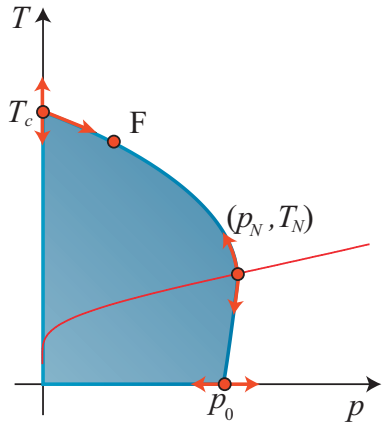}
\end{center}
\caption{Phase diagram of the $q$-state Potts gauge glass.
The notation is as in Figure \ref{paper_fig1}.
There is an extra critical point located at F along the phase boundary.
The arrows indicate the flows under the renormalization group.}
\label{fig:PGG}
\end{figure}

The schematic phase diagram of the Potts gauge glass in two dimensions is shown in
Figure~\ref{fig:PGG}. It differs from the left panel of Figure~\ref{paper_fig1} by the
existence of an extra critical point F on the phase boundary for $q>2$. The critical properties
at F have been shown numerically \cite{Jacobsen2002} to coincide with those of the
$q$-state random-bond Potts model \cite{Jacobsen1997,Jacobsen1998}.
Indeed, in a perturbative treatment the interaction between replicas involves the
product of the local energy densities, and this coupling is relevant for $q>2$ and
marginally irrelevant for $q=2$. For $q>2$, the renormalization group (RG) flow along the phase boundary is therefore from
the pure Potts point towards F, and from the Nishimori point towards either F or the
zero-temperature point at $p_0$.

The gauge symmetry is now obtained along the curve \cite{Nishimori1983,Jacobsen2002}
\begin{equation}
 {\rm e}^{K_{\rm Potts}} = \frac{p}{1-(q-1)p}
 \label{NlinePotts}
\end{equation}
which generalizes the Nishimori line (\ref{Nishimoriline}) to the situation $q > 2$.

The perturbative calculation for the random-bond Ising model \cite{Domany1979} cannot be straightforwardly generalized to the $q$-state case, since the application of the perturbation theory relies on the irrelevance of the random-bond disorder. 
This hypothesis is not fulfilled because of the RG flow from $T_c$ to point F, shown in Figure~\ref{fig:PGG}. 
Below we shall compute the correct slope to good numerical precision (see Table~\ref{table3}) and find it to be more accurate than the existing numerical estimation \cite{Jacobsen2002}.

\section{Duality with real-space renormalization}
\label{sec3}

In this section we review the duality analysis with real-space renormalization
\cite{Ohzeki2008,Ohzeki2009a}. We shall see later, in section~\ref{sec5}, how to improve
this method and overcome its shortcomings.

\subsection{Standard duality analysis}

The partition function of many statistical models, including the Ising and Potts models,
can be written as
\begin{equation}
 Z(x_0,x_1,\cdots) = \sum_{\phi_i = 0}^{q-1} \prod_{\langle ij \rangle} x_{\phi_i - \phi_j} \,,
\end{equation}
where $x_{\phi}$ is the local part of the Boltzmann factor, often referred to as the edge Boltzmann factor,
and $\phi$ denotes the discrete spin variable. For instance, for the Ising model ($q=2$) the orginal
spin reads $S_i = \cos \pi \phi_i$ and the edge weight of the random-bond model is
then $x_{\phi} = \exp(K\tau_{ij}\cos \pi \phi)$.

The standard duality transformation can be generalized by use of the discrete Fourier
transformation \cite{Wu1976}. Writing the dual edge Boltzmann factor as
\begin{equation}
 x_{k}^* = \frac{1}{\sqrt{q}} \sum_{\phi} x_{\phi} {\rm e}^{{\rm i} k \phi}.\label{dualBF}
\end{equation}
One can relate the original partition function to another one with dual edge Boltzmann factors as
\begin{equation}
Z(x_0,x_1,\cdots) = Z^*(x^*_0,x^*_1,\cdots) \,.
\end{equation}
In the partition function on the right-hand side, the edge-Boltzmann factor is replaced by its
dual, given in eq.~(\ref{dualBF}). 
We extract the principal Boltzmann factors $x_0$ and $x_0^*$ to obtain
\begin{equation}
\left(x_0\right)^{N_B} z(u_1,\cdots) = \left(x_0^*\right)^{N_B} z^*(u^*_1,\cdots) \,, \label{Zd}
\end{equation}
where $N_B$ is the number of the bonds (edges), and $z$ and $z^*$ are the normalized
partition functions. A duality relation can be obtained 
if we find that the normalized partition functions take the same functional form of the relative
Boltzmann factors $u_k = x_k/x_0$ and $u^*_k = x_k^*/x_0^*$. If this is the case,
we can rewrite $u^*_k$ in terms of $u_k$, but with different parameters (e.g., the dual coupling
$K^*$). For instance, for the pure Ising model this leads to the well-known \cite{Kramers1941}
duality relation $\exp(-2K^*) = \tanh K$.

Under the standard assumption of uniqueness of the critical point, the latter is then identified
as the fixed point of the duality relation. For the pure Ising model this leads to
eq.~(\ref{Isingselfdual}).
In particular, the principal Boltzmann factors $x_0$ and $x_0^*$ coincide with each other at
the fixed point of the duality relation.

\subsection{Duality with replica method}

The generalization of the standard duality transformation to random spin systems can be
obtained straightforwardly by considering the replicated partition function
\cite{Nishimori2002,Mailard2003}.
The edge Boltzmann factor is then given by $x_{\{\phi^a\}} = [ \exp(K\tau_{ij}\cos\pi \phi_a) ]$,
taking again the Ising model as an example.
The dual edge Boltzmann factor is calculated by multiple discrete Fourier transformations,
leading to the following relationship between the replicated partition functions:
\begin{equation}
 Z_n(x_0,x_1,\cdots) = Z_n^*(x^*_0,x^*_1,\cdots) \,, \label{Zr0}
\end{equation}
where $Z_n = [Z^n]$ denotes the configurational average $[\cdots]$ over the quenched random
variables $J_{ij}$, namely $\tau_{ij}$. We can index the variables
$x_k$ so that $k$ stands for the number of replicas in which the two nearest neighbour spins
take different values.

In general, for a finite number $n$ of replicas, a common functional form of $Z$ and $Z^*$ can only be
obtained at the price of
introducing edge interactions between an arbitrary number of replicas, even though physically
it would suffice to have pairwise interactions. This implies that the selfduality condition will
not be sufficient to fix all the coupling constants, or, in other words, the dimension of the selfdual
manifold is higher than zero.%
\footnote{This situation has been investigated in details for the case of replicated Potts
models \cite{Jacobsen1999,Jacobsen2000,Jacobsen2002a}.}
It is then not obvious how one can obtain a duality relation between the normalized partition
functions in the replica limit $n \to 0$.
On the other hand, even in such a situation it is of course easy to
calculate the original and dual edge-Boltzmann factors \cite{Nishimori2007}; see eq.~(\ref{dualBF}).

The replica approach to duality assumes that a good approximation to the fixed point of
the duality relation, directly in the replica limit, is given by the following single equation  %
\begin{equation}
x_0 =x_0^* \,. \label{Aeq0}
\end{equation}
The principal Boltzmann factor on the left-hand side is simply the edge Boltzmann factor with edge
spins parallel in all replicas, while the right-hand side is given by the direct manipulation of the
multiple discrete Fourier transformations.

It was shown in \cite{Nishimori2002,Mailard2003} that for the random-bond Ising model eq.~(\ref{Aeq0}) 
indeed leads to a rather precise approximate value of the critical point on the Nishimori line,
$p_N \simeq 0.110028$, which was initially conjectured to be exact \cite{Mailard2003}.
However, it marginally disgrees with the numerical result $p_N = 0.10919(7)$ \cite{Hasenbusch2008}.
Fortunately, it turns out that eq.~(\ref{Aeq0}) is only the first in a series of
similar approximations in which the precision is systematically improved 
\cite{Ohzeki2008,Ohzeki2009a}.

\subsection{Real-space renormalization}

\begin{figure}[tb]
\begin{center}
\includegraphics[width=120mm]{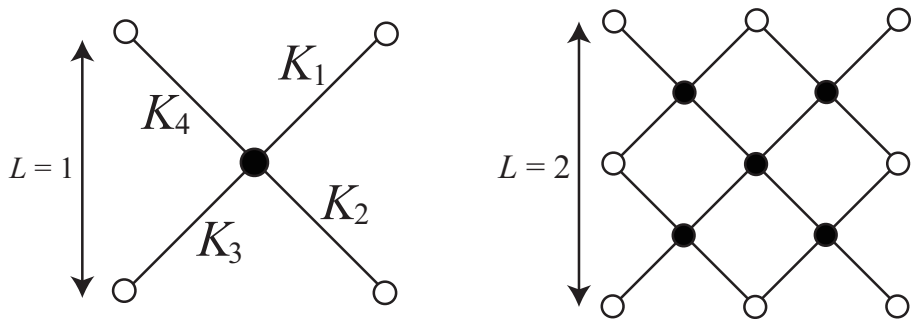}
\end{center}
\caption{The left panel depicts the simplest basis for the duality analysis with real-space renormalization, consisting of four
bonds and henceforth referred to as the basis of size $L=1$.
The right panel shows the second basis for the duality with real-space renormalization:
a $16$-bond cluster henceforth referred to as the $L=2$ basis.}

\label{fig4}
\end{figure}

One of the present authors has proposed an improvement of the above replica computation
which relies on
the systematic introduction of real-space renormalization \cite{Ohzeki2008,Ohzeki2009a}.
In this new approach one now uniformly sums over a part of the degrees of freedom
in both of the partition functions in eq.~(\ref{Zr0}), which we rewrite as
\begin{equation}
Z_n(x^{({\rm B})}_0,x^{({\rm B})}_1,\cdots) = Z_n^{*}(x^{*({\rm B})}_0,x^{*({\rm B})}_1,\cdots) \,.
\label{ZnRSrenorm}
\end{equation}
The summation is over the spins inside some finite portion of the lattice $B$, called the
basis, whose linear size $L$ characterizes the degree
of the real-space renormalization. 
The simplest bases with $L=1$ and $L=2$ are depicted in Figure \ref{fig4}. Here, the
white circles denote the $4L$ spins that are fixed in a given reference state ($S_i = +1$ for Ising),
while the black circles are the $N_L = 2L^2 - 2L + 1$ internal spins which are summed over in the
real-space renormalization. The standard duality analysis is reproduced when $L=1$.%
\footnote{
The standard duality analysis \cite{Nishimori2002,Mailard2003} involves a basis with only one bond,
and hence does not correspond to a definite value of $L$ in our notation.
}

The edge Boltzmann factor includes many-body interactions which the real-space renormalization
generates between the remaining degrees of freedom. This has been emphasized by the
superscript on the edge Boltzmann factors $x_k^{({\rm B})}$ in eq.~(\ref{ZnRSrenorm}).
The assumption is now that the critical surface of the quenched random system is the
solution of
\begin{equation}
 x^{({\rm B})}_0 =x_0^{*({\rm B})} \,. \label{Aeq1}
\end{equation}
As before the renormalized principal Boltzmann factor $x^{({\rm B})}_0$ corresponds to all spins
being parallel, and $x^{*({\rm B})}_0$ is defined by the discrete Fourier transforms.

It is an important property of the method that if the simplest condition (\ref{Aeq0}) already
captures the exact
fixed point of the duality relation, it does not change even after real-space renormalization.
Indeed for the Ising model and the Potts model without any disorder (thus without replicas),
the solution of eq.~(\ref{Aeq0}) is also a solution of eq.~(\ref{Aeq1}) for any size $L$ of the basis.

Conversely, when eq.~(\ref{Aeq1}) exhibits a dependence on $L$ we expect it to provide
approximations to the critical point whose accuracy increases systematically with $L$.
This expectation has been verified for spin glasses \cite{Ohzeki2008,Ohzeki2009a}.
We further expect the approximations to tend to the exact result in the limit $L \to \infty$,
but obviously explicit computations are limited by the value of $L$ for which we are able
to carry out the summation over the $q^{N_L}$ possible values of the internal spins.

It has also been verified that eq.~(\ref{Aeq1}) can yield the well-known percolation threshold
manifold of the inhomogeneous bond-percolation problem \cite{Ohzeki2013a}.

\section{Critical polynomial}
\label{sec4}

The other author and Scullard have defined another independent extension of the duality
method \cite{Jacobsen2012,Jacobsen2013}, motivated initially by the desire to obtain critical 
points for pure statistical models, such as the $q$-state Potts model,
to a precision largely exceeding that of available numerical simulations \cite{Jacobsen2014}.
Their construction is known as the critical polynomial, and we shall begin by reviewing its definition and some
of its main properties. Surprisingly many of these are very similar to those of the
duality method with real-space renormalization. We shall therefore continue by bringing
the critical polynomial into a form (with $q$ integer) that allows us to compare the two constructions.
This will ultimately lead to a modification of the real-space renormalization approach to
quenched random systems that we shall derive in section~\ref{sec5} and
apply to spin glass systems in section~\ref{sec6}.

\subsection{Definition and main properties}

Given any connected graph $G=(V,E)$ with vertex set $V$ and edge
set $E$, the partition function $Z$ of the $q$-state Potts model can be defined
in the so-called Fortuin-Kasteleyn (FK) representation as \cite{FK1972}
\begin{equation}
 Z = \sum_{A \subseteq E} v^{|A|} q^{k(A)} \,,
 \label{FK_repr}
\end{equation}
where $|A|$ is the number of edges in the subset $A$, and $k(A)$
the number of connected components (including isolated vertices)
in the subgraph $G_A = (V,A)$.
The temperature variable is written $v = {\rm e}^K_{\rm Potts} - 1$,
where $K_{\rm Potts}$ is the reduced interaction
energy between adjacent $q$-component spins.
It is an advantage of the representation (\ref{FK_repr}) that $q$
can formally be allowed to take arbitrary real values.

Suppose we compute the partition function of a finite portion $B$, of size $L \times L$,
of a planar lattice with periodic boundary conditions in both lattice directions.
As before, $B$ is referred to as the basis and it is illustrated in Figure \ref{fig4}, the only
difference being that opposed pairs of white circles are identified by the periodic
boundary conditions. In particular, all the spins (black or white) are now internal spins
to be summed over, and their number is $N_L = 2 L^2$. The terms in eq.~(\ref{FK_repr})
can now be distinguished by the homotopy properties of $G_A$ on the torus \cite{Jacobsen2013},
or, more precisely, by those of its connected components (commonly known as FK clusters).
We shall need to distinguish three classes of terms:
\begin{enumerate}
 \item Those where all FK clusters are homotopic to a point are referred to as 0D configurations.
 \item Those where all cluster boundaries are homotopic to a point, but there exists a (unique)
 FK cluster that is non-contractible. We call these 2D
 configurations; note that they are nothing but the duals of the 0D configurations.
 \item The remaining terms are referred to as 1D configurations. They are characterized by
 the existence of both a non-contractible cluster and a non-contractible cluster boundary.
\end{enumerate}
An operational method for distinguishing 0D, 1D and 2D configurations by the computation of
an Euler characteristics has been given in \cite{Jacobsen2013,Jacobsen2014}. The corresponding
partial sums of (\ref{FK_repr}) are denoted $Z_{\rm 0D}$, $Z_{\rm 1D}$ and $Z_{\rm 2D}$,
and we have obviously $Z = Z_{\rm 0D} + Z_{\rm 1D} + Z_{\rm 2D}$. The graph polynomial
$P_B(q,v)$ introduced in \cite{Jacobsen2012} can then be rewritten \cite{Jacobsen2013} simply as
\begin{equation}
 P_B(q,v) = Z_{\rm 2D} - q Z_{\rm 0D} \,.
\end{equation}

The real roots of $P_B(q,v)$ enjoy a number of properties that closely resemble those
of the duality method with real-space renormalization. In particular, the definition with respect
to a basis $B$ whose size $L$ can be increased is quite similar.
It was found in \cite{Jacobsen2012,Jacobsen2013,Jacobsen2014}
for a wide variety of lattices, including all Archimedian lattices, their duals and their medials,
that for exactly solvable cases $P_B(q,v)$ factorizes over the integers, shedding a small factor that
is independent of $L$ and one (or more) of whose zeros provides the exact critical point(s).
Moreover, when this factorization does not take place, the relevant root of $P_B(q,v)$ was found
to converge very fast towards the critical point $v_{\rm c}$ upon increasing $L$. In \cite{Jacobsen2014} this property was used to determine $v_{\rm c}$ to 12 or 13 digit precision
in the most favourable cases.

Finally, the graph polynomial method can be applied to cases where the basis supports
inhomogeneous couplings. In particular, the case of the square lattice with checkerboard
interactions was investigated in \cite{Jacobsen2012} and the relation of a more general
class of inhomogeneous bow-tie lattices to quantum integrability was elucidated in \cite{Scullard2013}.
Below we shall take the next natural step of applying the graph polynomial method to
quenched random systems.

\subsection{Case of integer $q$}

It is obvious that the graph polynomial method differs from the approach of section~\ref{sec3} due to
the different boundary conditions. To make a more detailed comparison, we have evaluated
$P_B(q,\{v\})$ for a variety of inhomogeneous lattices and bases $B$ for several integer
values of $q=2,3,\ldots$, setting $v_{ij} = w_{ij} - 1$ and expanding the result as a polynomial
in the edge Boltzmann factors $w_{ij} = {\rm e}^{K_{ij}}$ appearing in the original (integer) spin
formulation of the Potts model with Hamiltonian
\begin{equation}
 H = - \sum_{\langle ij \rangle} K_{ij} \delta_{S_i,S_j}
\end{equation}
and $S_i = 0,1,\ldots,q-1$.

In this way we have established%
\footnote{One of us (JLJ) thanks C.R.~Scullard for discussions about this calculation.}
that for the Ising model
\begin{equation}
 P_B(2,v) = Z_{++} - Z_{+-} - Z_{-+} - Z_{--} \,,
\end{equation}
where $Z_{\tau_x,\tau_y}$ denotes the partition function on the corresponding basis $B$,
the Ising spins being endowed with boundary conditions in the horizontal direction that are
periodic ($\tau_x = +$) or antiperiodic ($\tau_x = -$), and $\tau_y$ similarly denoting
the boundary conditions in the vertical direction.

This result extends to the Potts model with integer $q$ as follows:
\begin{equation}
 P_B(q,v) = q Z_{0,0} - \sum_{\tau_x,\tau_y = 0}^{q-1} Z_{\tau_x,\tau_y} \,,
 \label{PBqinteger}
\end{equation}
where the twists $(\tau_x,\tau_y)$ of the boundary conditions signify that a pair
of neighbouring spins, $\sigma_i$ and $\sigma_j$, crossing the horizontal periodic boundary
condition are subject to a twisted interaction term $\delta_q(S_i-S_j+\tau_x)$, cf.~eq.~(\ref{H_PGG}),
and similarly for the vertical twist $\tau_y$.

In the next section we shall see that applying these sums over twisted periodic boundary conditions
to the spin glass problem provides estimates for the critical manifold with much better convergence
properties than the original approach of duality with real-space renormalization (see section~\ref{sec3})
in which the spins of the basis were coupled to fixed boundary spins.

\section{Characteristic equation for quenched random systems}
\label{sec5}

Although the previous duality analysis with real-space renormalization has been
succesfully applied to random spin systems \cite{Ohzeki2008,Ohzeki2009a}
as well as to pure bond percolation problems \cite{Ohzeki2013a}, it contains
some ambiguity in the choice of boundary conditions. As shown in Figure~\ref{fig4}
the renormalized principle Boltzmann factor was computed with all boundary spins
of the basis $B$ fixed and parallel, but this convention was rather {\em ad hoc} and
simply taken by inspiration with the original replicated duality analysis
\cite{Nishimori2002,Mailard2003}.

It is thus perfectly reasonable to impose other boundary conditions for the computation of the
renormalized principal Boltzmann factors $x_0^{({\rm B})}$ and $x_0^{*({\rm B})}$
to be inserted in the characteristic equation
(\ref{Aeq1}). The superior precision of the critical polynomial method when applied, e.g.,
to the bond percolation threshold on the kagome lattice \cite{Jacobsen2012,Ohzeki2013a},
gives solid guidance that it might be profitable to adopt its inherent boundary conditions
on $B$, as discussed in section~\ref{sec3}.

Obviously the consequences of making this choice
for quenched random systems should then be carefully compared with the previous results
derived by the standard duality and the duality analysis with real-space renormalization.

\subsection{Change of the boundary conditions}

Motivated by these remarks, we hence change the boundary conditions. Rather than
having a set of fixed parallel boundary spins (white circles in Figure~\ref{fig4})
we henceforth impose periodic conditions in both lattice directions when we compute
the renormalized principal Boltzmann factor. To make this choice explicit we henceforth
denote it as $x_0^{({\rm PB})}$ and its dual as $x_0^{*({\rm PB})}$. The white circles
in Figure~\ref{fig4} are then considered also as internal spins, and opposing pairs of
white circles are identified by the periodic boundary conditions.

For instance, for the simplest
case $L=1$ there are now two spins to be summed over: $S_0$ (the black circle) and
$S_1$ (the white circles, all of which have been identified).
Because of this identification, and the $S_q$ symmetry of the interactions, the
change of boundary condition actually only provides an overall factor in this case,
and the result of applying eq.~(\ref{Aeq1}) is hence unchanged. This is apparent in the
tables below and also, for example, in the agreement between the methods of sections
\ref{sec3} and \ref{sec4} in reproducing the Wu conjecture \cite{Wu1979} for the bond percolation
threshold on the kagome lattice when $L=1$ \cite{Jacobsen2012,Ohzeki2013a}.
In other words, the proposed method is consistent with the original duality analysis
without real-space renormalization \cite{Nishimori2002,Mailard2003}.

We now consider in more detail the case $L=1$ of the replicated Ising model with the
new boundary conditions. The renormalized principal Boltzmann factor reads
\begin{eqnarray}\nonumber
x_0^{({\rm PB})} &=& \left[ \left(\sum_{S_0,S_1} \exp\left\{\left(K_1+K_2+K_3+K_4\right)S_0S_1\right\}\right)^n\right]
\\ &=& 4^n\cosh^n \left(K_1+K_2+K_3+K_4\right) \,,
\end{eqnarray}
where we have set $K_i = K \tau_i$. In the replica limit $n \to 0$ this reduces to
\begin{equation}
x^{({\rm PB})}_0 = 1 + n \left[ \log Z^{({\rm PB})}_{++} \right] \,,
\end{equation}
where $Z^{({\rm PB})}_{++} = 4 \cosh(K_1+K_2+K_3+K_4)$ represents the partition function
of the inhomogeneous Ising model on the dual graph of the $L=1$ basis, and the subscripts
indicate that periodic boundary conditions have been applied in both lattice directions.  

The dual principal Boltzmann factor, still with doubly periodic boundary conditions, meanwhile
becomes
\begin{eqnarray}
x^{*({\rm PB})}_0 &=& \frac{1}{4^n} \left(\sum_{S_0,S_1} \prod_{i=1}^4\left\{ \exp\left(K_i\right) + S_0S_1 \exp\left(-K_i\right)\right\}\right)^n  \\
&=& 2^n \left(\left\{ \prod_{i=1}^4 \cosh K_i + \prod_{i=1}^4 \sinh K_i \right\}\right)^n \nonumber \\
&=& 2^n \left[ \cosh(K_1+K_2+K_3+K_4) + \cosh(K_1+K_2-K_3-K_4) \right. \nonumber \\
& & + \left. \cosh(K_1-K_2+K_3-K_4) + \cosh(K+1-K_2-K_3+K_4) \right]^n \,. \nonumber
\end{eqnarray}
Taking the replica limit leads to
\begin{equation}
x_0^{*({\rm PB})} = 1+n \left( [\log Z^{({\rm PB})}] - \log 2\right) \,,
\end{equation}
where now
$Z^{({\rm PB})}=Z^{({\rm PB})}_{++} + Z^{({\rm PB})}_{+-} + Z^{({\rm PB})}_{-+} + Z^{({\rm PB})}_{--}$,
and $Z^{({\rm PB})}_{\tau_x,\tau_y}$ denotes the partition function of the inhomogeneous Ising model
on the dual graph with the boundary conditions in either direction ($i=x,y$) specified by $\tau_i$ as
being periodic ($\tau_i = +1)$ or antiperiodic $(\tau_i = -1)$.
In other words, the periodic boundary conditions on the spins implies a sum over all different
boundary conditions (periodic and antiperiodic) for the dual principal Boltzmann factor.

Finally, the criterion (\ref{Aeq1}) for estimating the critical points of spin glasses becomes
\begin{equation}
 x_0^{({\rm PB})} = x_0^{*({\rm PB})}
 \label{AeqPB}
\end{equation}
in the present notation, and reads explicitly
\begin{equation}
\left[ \log Z^{({\rm PB})} \right]  - \left[ \log Z^{({\rm PB})}_{++}\right]  = \log 2 \,. \label{Aeq2}
\end{equation}
It is shown in \ref{appA} than even for larger bases, $L \ge 1$, the application of eq.~(\ref{AeqPB})
to the renormalized principal Boltzmann factors computed with periodic boundary conditions on the
basis still produces the same expression as in eq.~(\ref{Aeq2}). This means that the asymptotic
analysis for larger $L$ can be performed straightforwardly, in analogy with \cite{Ohzeki2009b}.

\subsection{Relation to quantum error correction}

The hypothesis that eq.~(\ref{Aeq2}) identifies the location of the critical points of spin glasses
is also reasonable from a couple of other ponts of view.
Twisted boundary conditions are indeed utilized for investigation of the rigidity of the ordered state \cite{Hukushima1999}.

Moreover, in the realm of quantum error correction, twisted boundary conditions serve
to detect the phase transition of the error correctability  \cite{Dennis2002}.
In that context, the quantum state is encoded as a many-body system with qubits placed
on each bond on the square lattice.
We encode the distinguished two-bit state by imposing periodic boundary conditions in both directions.
The ratio $Z_{++}/Z$ stands for the likelihood to correctly infer the encoded quantum state.
On the other hand, $Z_{+-}/Z$, $Z_{-+}/Z$, and $Z_{--}/Z$ represent the likelihood of the different undesired quantum states.

If the possibility of the undesired states is suppressed, meaning that the qubits are not prone
to errors, we have $Z_{++}/Z \to 1$.
This can be expressed as $\log Z - \log Z_{++} = 0$.
In the other extreme, when the true quantum state cannot be inferred due to the preponderance
of errors, $Z_{++}/Z$ becomes close to $1/4$, meaning that $\log Z - \log Z_{++} = 2\log2$.
Therefore the obtained eq.~(\ref{Aeq2}) states that the middle point between the above
extreme cases should be the critical point, in the limit of a large number of qubits, namely
\begin{equation}
\log Z  - \log Z_{++}  = \log 2 \,. \label{Aeq22}
\end{equation}
We emphasize that the partition function appearing in the above equation is for the whole system
that holds the quantum state.
On the other hand, our expression (\ref{Aeq2}) is for the small subsystem defining the basis.

\subsection{Generalization to the Potts model}

As shown in \ref{appA}, it is straightforward to use the periodic boundary conditions to
derive the renormalized principal Boltzmann factors in the more general case of
the $q$-state Potts model. The criterion (\ref{AeqPB}) then becomes
\begin{equation}
\left[ \log \left( \sum_{\tau_x,\tau_y=0}^{q-1} Z^{({\rm PB})}_{\tau_x,\tau_y}\right) \right]
- \left[ \log Z^{({\rm PB})}_{00} \right] = \log q \,.
 \label{generalPottsresult}
\end{equation}
Here $(\tau_x,\tau_y)$ denotes the twist of the boundary condition in both of the periodic directions.
If we restrict ourselves to the case of the Potts model without any disorder and do not use the
replica method, we obtain 
\begin{equation}
\left( \sum_{\tau_x,\tau_y=0}^{q-1} Z^{({\rm PB})}_{\tau_x,\tau_y}\right) - q Z^{({\rm PB})}_{00} = 0 \,.
\label{criterionpurepotts}
\end{equation}
By eq.~(\ref{PBqinteger}) this is precisely the same result as the criterion $P_B(q,v)=0$
obtained from the critical polynomial approach. We have thus proved that, once periodic
boundary conditions are imposed on the underlying basis $B$, the two methods
(duality with real-space renormalization, and the critical polynomial) are equivalent.

Working the other way around, if we apply the replica trick to the critical polynomial
result we recover eq.~(\ref{generalPottsresult}).

We note that the criterion (\ref{criterionpurepotts}) for the inhomogeneous Ising model ($q=2$),
\begin{equation}
Z^{({\rm PB})}_{+-}  + Z^{({\rm PB})}_{-+}  + Z^{({\rm PB})}_{--} - Z^{({\rm PB})}_{++} = 0 \,,
\end{equation}
has previously been obtained by direct manipulation of the standard duality transformation
in \cite{Bugrii1996}.

\section{Applications to spin glasses}
\label{sec6}

The limiting factor in turning eq.~(\ref{generalPottsresult}) into a practical means of computing
the phase diagram of spin glasses is that the disorder average $[\cdots]$ must be performed
over all realizations of the $N_L$ random coupling constants in the basis $B$. For the $q$-state
Potts gauge glass this implies a sum over $q^{N_L}$ terms, and in the special case of the
random-bond Ising model we have $q=2$.

For the basis shown in Figure~\ref{fig4}, where the square lattice is oriented diagonally,
there are $2L^2$ spins and $N_L = 4 L^2$ bonds. We shall find it useful in the following to also
consider bases where the lattice has a straight orientation, with $L^2$ spins and $N_L = 2 L^2$
bonds. These two types of bases will be referred to as ``diagonal'' and ``straight'' respectively.
The straight basis has the advantage that one can access a larger number of different values
of $L$.

In addition, for each disorder realization the various partition functions
$Z_{\tau_x,\tau_y}^{({\rm PB})}$ must be computed. This is straightforward
to do by transfer matrix techniques with sparse matrix factorization \cite{Jacobsen1998}
in a time of the order $N_L \times q^L$.
In the critical polynomial studies without disorder, diagonal bases of sizes up
to $L=7$ were considered \cite{Jacobsen2014}, but obviously
we must now content ourselves with smaller bases, since in addition to computing
the $Z_{\tau_x,\tau_y}^{({\rm PB})}$ we will also have to average over the disorder.

\subsection{Phase diagram of the random-bond Ising model}

\begin{figure}[tb]
\begin{center}
\includegraphics[width=120mm]{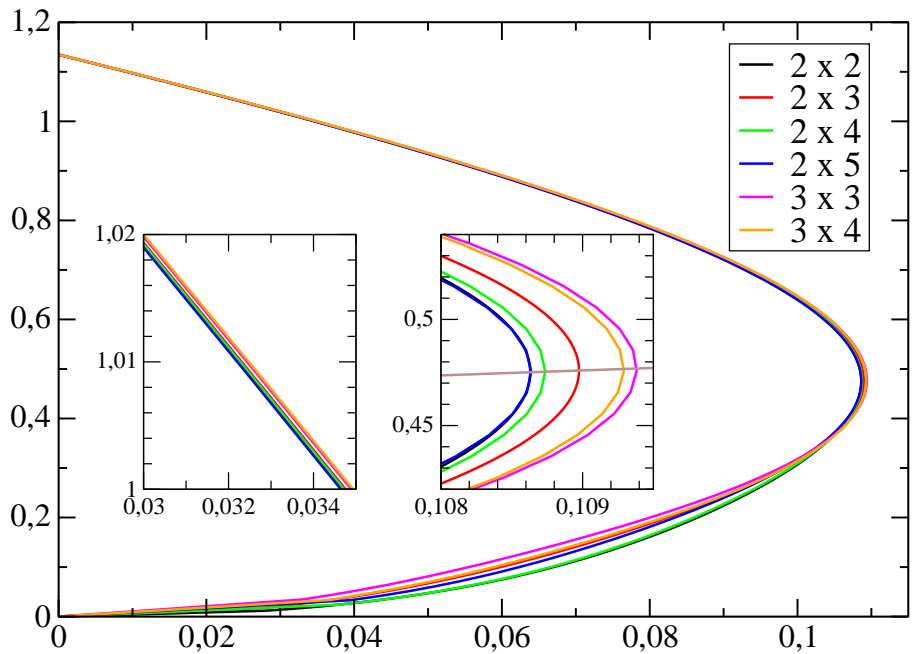}
\end{center}
\caption{Phase diagram of the random-bond Ising model in the $(p,T)$ plane for various choices of rectangular
bases of size $L \times M$. The two insets show zooms on selected regions of the phase diagram.
Left inset: In the high-temperature region the finite-size effects are almost negligible.
Right inset: At the intersection with the Nishimori line (shown in brown) all the curves have
a vertical tangent.}
\label{phase-diagram}
\end{figure}

The phase diagram for the random-bond Ising model with several choices of
straight bases is shown in Figure~\ref{phase-diagram}. In addition to square-shaped
bases (with equal height and width) we have also considered $L \times M$
rectangular-shaped bases, invariably with the straight orientation.

The size-dependence in the high-temperature region is very slight, meaning that the obtained
phase boundary is essentially exact. The same could be said about Figure~\ref{paper_fig2},
so below we compare the present approach to the previous one \cite{Ohzeki2009a} in a more
quantitative way. In particular we shall focus on the precise determination of the Nishimori
point $p_N$ and the slope of the phase boundary around the pure Ising point.

As in Figure~\ref{paper_fig2}, the low-temperature region is subject to larger finite-size
corrections. Since all curves go though the origin, $(p,T) = (0,0)$, one may first think that
there is a discrepancy with the fact (shown in Figure~\ref{paper_fig1}) that there is a
zero-temperature critical point at finite $p_0$. Below we shall however show analytically
that the slope at the origin tends to zero when $L \to \infty$, which guarantees the agreement
with Figure~\ref{paper_fig1}.

All the curves have a vertical tangent at the Nishimori point, in agreement with an
argument presented in \cite{Ohzeki2009a}.

\subsection{Slope at the pure Ising point}
\label{sec:extrapolslope}

We now start applying the modified version of the theory, namely the duality analysis with real-space
renormalization under periodic boundary conditions, in a more quantitative way. To this end, we first
study its convergence properties upon estimating the slope of the phase boundary at the pure Ising
(or Onsager) point.

Consider the expansion around the Ising point to first order in $p = \Delta p$ (with $\Delta p \ll 1$) of eq.~(\ref{Aeq2})
\cite{Ohzeki2011a}. We obtain the following relation for the slope at $(p,T) = (0,T_{\rm c})$:
\begin{eqnarray}\nonumber
\alpha \equiv \frac{1}{T_c}\frac{\Delta T}{\Delta p} &=& T_c \left( \frac{1}{Z^{\rm F}}\frac{dZ^{\rm F}}{dK} - \frac{1}{Z^{\rm F}_{++}}\frac{dZ^{\rm F}_{++}}{dK} \right)^{-1}\\
& & \quad \times \left\{\sum \left( \log Z^{{\rm AF}_1}(K_c) -\log Z^{{\rm AF}_1}_{++} (K_c) \right) \right\} \,, \label{DTc}
\end{eqnarray}
where for simplicity we have omitted the superscript $({\rm PB})$.
Instead, the superscript F denotes the partition function only with ferromagnetic interactions
and ${\rm AF}_x$ stands for that with $x$ antiferromagnetic interactions.
The summation runs over all configurations having $x$ antiferromagnetic interactions.
We recall that according to the perturbative argument \cite{Domany1979},
the exact value of the left-hand side of eq.~(\ref{DTc}) in the thermodynamic limit
is given by eq.~(\ref{slope}) as $\alpha = 3.209112647 \cdots$.

\begin{table}
\begin{center}
\caption{Estimations of the slope $\alpha$ from eq.~(\ref{DTc}) in the $\pm J$ random-bond Ising model, using diagonal bases of size $L$.
The middle column lists the values given by the original duality analysis \cite{Ohzeki2009a} using
fixed boundary conditions, and the right column represents our modified version using
periodic boundary conditions.}
\label{table1}
\begin{tabular}{cll}
$L$ & $\alpha^{\rm (F)}$ & $\alpha^{\rm (PB)}$ \\
\hline
1  & $3.33658107$ & $3.33658107$\\
2  & $3.31271593$ & $3.23692555$\\
3  & $3.29351615$ & $3.22106163$\\
4  & $3.28160520$ & $3.21573448$\\
5  & $3.27362975$ & $3.21331854$\\
6  & $3.26791553$ & $3.21202079$\\
7  & $3.26360719$ & $3.21124352$\\
8  & $3.26023098$ & $3.21074121$\\
9  & $3.25750489$ & $3.21039784$\\
\hline
$\infty$ & & $3.20911265$
\end{tabular}
\end{center}
\end{table}

\begin{figure}[tb]
\begin{center}
\includegraphics[width=80mm]{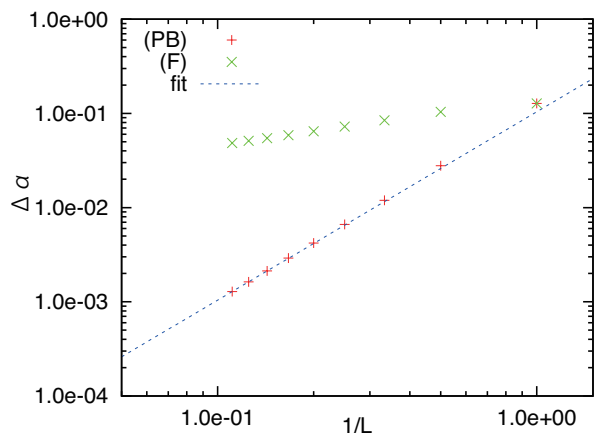}
\end{center}
\caption{Difference $\Delta \alpha$ between our estimations for the slope $\alpha$ of the phase boundary in the $\pm J$ Ising model and the exact $L \to \infty$ result by Domany \cite{Domany1979}.
The green crosses show the previous results for $\Delta \alpha$, obtained by duality with real-space renormalization as in Ref.~\cite{Ohzeki2009a}.
The red plusses our new estimations from eq.~(\ref{DTc}).
The blue line represents a fit of the form $\Delta \alpha \simeq aL^{-\omega}$.
}
\label{slope-fig}
\end{figure}

We list our estimations on the slope value obtained from both the original \cite{Ohzeki2009a} 
and our modified duality analysis in Table \ref{table1}.
The modified version with periodic boundary conditions exhibits notably faster convergence
to the exact result, as shown in Figure~\ref{slope-fig}.

Note that with a single antiferromagnetic interaction and periodic boundary conditions, each term
in the sum over the $N_L$ such configurations can only take two distinct values,
depend on the orientation of the frustrated bond. We have checked by explicit computations that
this is indeed the case.

To examine the convergence rate of the data in Table~\ref{table1}, we have investigated the difference
$\Delta \alpha$ between the data and the exact result (\ref{DTc}). For each $L_0=1,2,\ldots,L_{\rm max}-1$,
we have first fitted the data points $\Delta \alpha$ with size $L \ge L_0$ to the pure power law $a(L_0) L^{-\omega(L_0)}$. 
The exponents $\omega(L_0)$ thus obtained are then fitted to a second-order polynomial in $1/L_0$ in order to assess their residual finite-size dependence. In the case of periodic
boundary conditions, iteration of this procedure
leads to the result
\begin{equation}
 \Delta \alpha^{\rm (PB)} \propto L^{-\omega^{\rm (PB)}} \quad \mbox{with} \quad  \omega^{\rm (PB)} = 2.000 (1) \,.
\end{equation}
We conjecture the exact value of this exponent to be
\begin{equation}
 \omega^{\rm (PB)} = 2 \,.
 \label{conjFSSexp}
\end{equation}
For fixed boundary conditions \cite{Ohzeki2009a} the finite-size dependence of the initial fits is
much more pronounced. The final polynomial fit is however still very good and leads to the estimate
\begin{equation}
 \omega^{\rm (F)} = 0.3114(3) \,.
 \label{FomegaRBIM}
\end{equation}

The higher value of $\omega^{\rm (PB)}$ entails a considerable improvement over \cite{Ohzeki2009a}.
In the absense of disorder, this exponent may be even higher. For instance, in the application of the critical
polynomial method to bond percolation it was found that $\omega \simeq 6.35$ for several
different lattices \cite{Jacobsen2014,Jacobsen2015}.

We have also applied the modified duality analysis to the bond-diluted Ising model.
Estimates for the slope value are shown in Table \ref{table2}.
We again confirm the faster convergence to the exact result (\ref{dilute-slope}) found from
the perturbation theory \cite{Domany1978}. 
In this case the finite-size exponent is given by
\begin{equation}
 \omega^{\rm (PB)} = 2.003(1)\,,
\end{equation}
and we conjecture that also in this case is the exact value given by (\ref{conjFSSexp}).
This agrees with the expectation that the exponents should satisfy universality.

As before, the fits with free boundary conditions exhibit a stronger
finite-size dependence. The final value of the exponent comes out as
$\omega^{({\rm F})} = 0.319(2)$. This is sufficiently close to (\ref{FomegaRBIM})
that we can conjecture that also those two exponents coincide.

\begin{table}
\begin{center}
\caption{Estimations of the slope $\alpha$ from eq.~(\ref{DTc}) in the bond-diluted Ising model, using diagonal bases of size $L$.
The middle column lists the values given by the original duality analysis under fixed boundary conditions \cite{Ohzeki2009a} and the right one represents the modified version using periodic boundary conditions.}\label{table2}
\begin{tabular}{ccc}
$L$ & $\alpha^{\rm (F)}$ & $\alpha^{\rm (PB)}$ \\
\hline
1  & $1.33780277$ &$1.33780277$\\
2  & $1.33626191$ &$1.33120928$\\
3  & $1.33499619$ &$1.33010247$\\
4  & $1.33420150$ &$1.32972714$\\
5  & $1.33366586$ &$1.32955630$\\
6  & $1.33328059$ &$1.32946438$\\
7  & $1.33298914$ &$1.32940927$\\
8  & $1.33276034$ &$1.32937363$\\
9  & $1.33257531$ &$1.32934926$ \\
\hline
$\infty$ & &        $1.32925798$
\end{tabular}
\end{center}
\end{table}
\begin{figure}[tb]
\begin{center}
\includegraphics[width=80mm]{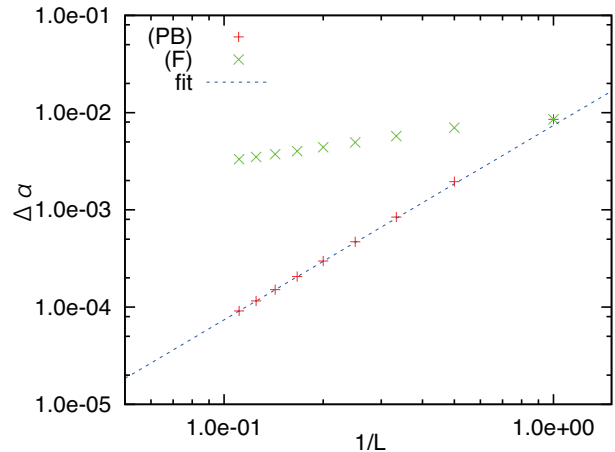}
\end{center}
\caption{ Difference $\Delta \alpha$ between our estimations for the slope of the phase boundary
in the bond-diluted Ising model and the exact $L \to \infty$ result \cite{Domany1978}.
The symbols used have the same meaning as in Figure \ref{slope}.
}
\label{slope_dil}
\end{figure}

\subsection{Nishimori point}

After these preliminaries, we are now ready to perform a high-precision estimation of the
Nishimori point in the $\pm J$ random-bond Ising model.

If we use a single bond as a basis, we can reproduce an approximate value of the Nishimori point as
$p_N = 0.110028$ \cite{Nishimori2002,Mailard2003}.
When we take a four-bond ($L=1$ diagonal) basis as in Figure \ref{fig4}, the application of
eq.~(\ref{generalPottsresult}) gives the estimate $p_N= 0.109275$, and by
increasing the size of the basis to $16$ bonds $(L=2)$ as in Figure \ref{fig4}
the estimate comes out as $p_N = 0.109097$.

This is slightly less than $p_N = 0.109178$ obtained
from the previous $L=2$ estimate by the duality with real-space renormalization (with
fixed boundary conditions) \cite{Ohzeki2009a}.
The location of the Nishimori point has been determined as $p_N = 0.10919(7)$
in a different numerical computation \cite{Hasenbusch2008}.

To take the computation to higher precision we now turn to straight bases with 
$N_L = 2 L^2$ bonds. The results for $p_N$ are given in Table~\ref{tabpN}.
The value for $L=1$ coincides with the single-bond computation \cite{Nishimori2002,Mailard2003}
as it should, since both of the two bonds connect the unique spin to itself.
The values for $L=1,2,3$ have been obtained by computing the full average over the
distribution of random bonds in eq.~(\ref{generalPottsresult}). For $L=4,5$ we have
truncated the distribution to the statistically most significant terms, as we now describe.

\begin{table}
\begin{center}
\caption{Determination of the Nishimori point of the $\pm J$ Ising model
from straight bases of size $L$.}
\label{tabpN}
\begin{tabular}{ll}
$L$ & $p_N$ \\
\hline
 1 & $0.110027864$ \\
 2 & $0.108630326$ \\
 3 & $0.109382432$ \\
 4 & $0.10912574 (6)$ \\
 5 & $0.109 (2)$ \\
\end{tabular}
\end{center}
\end{table}

\subsubsection{Truncation.}
\label{subsec:trunc}

Since the part of the phase diagram that is of interest to us is situated at $p \ll 1$
it makes sense to constrain the sum over the disorder realizations to configurations
having at most $k$ antiferromagnetic bonds. The corresponding estimates $p_N(k)$
can then be expected to converge rapidly in $k$.

\begin{table}
\begin{center}
\caption{Determination of $p_N$ for the straight $L=3$ basis by truncation.
The statistical sum is truncated to at most $k$ antiferromagnetic couplings.
The corresponding fraction of the total statistical weight at $p = 0.109$ is denoted $\rho(k)$.}
\label{tabtrunc3}
\begin{tabular}{rll}
$k$ & $p_N(k)$ & $\rho(k)$ \\
\hline
  3 & 0.140555390594 & 0.875004473 \\
  4 & 0.116481875139 & 0.960849527 \\
  5 & 0.111006230803 & 0.990254589 \\
  6 & 0.109687801570 & 0.998048635 \\
  7 & 0.109432720487 & 0.999683173 \\
  8 & 0.109388717859 & 0.999958118 \\
  9 & 0.109383115133 & 0.999995491 \\
10 & 0.109382487994 & 0.999999606 \\
11 & 0.109382436219 & 0.999999972 \\
12 & 0.109382432328 & 0.999999998 \\
\end{tabular}
\end{center}
\end{table}

We show the results of this truncation procedure for the straight $L=3$ basis in
Table~\ref{tabtrunc3}, for $3 \le k \le 12$. In this case we know the exact value
of $p_N = p_N(N_L)$, see Table~\ref{tabpN}, and we can see that the $k=12$
truncation captures correctly the first nine decimal digits.

Defining
\begin{equation}
 w_k(p) = \sum_{\ell = 0}^k p^\ell (1-p)^{N_L - \ell} {N_L \choose \ell}
\end{equation}
the fraction of the total statistical weight covered by the truncation at order $k$ is
\begin{equation}
 \rho(k) = w_k(p) / w_{N_L}(p) \,.
 \label{fracstatweight}
\end{equation}
The evaluation of $\rho(k)$ for $p = 0.109$ (a value close to the Nishimori point)
is shown in the last column of Table~\ref{tabtrunc3}. For instance, it is seen (for $k=6$)
that we can capture $99,8 \%$ of the statistical weight by summing over only $11,9 \%$ of the
disorder realizations.

Similar results for the straight $L=4$ basis with $5 \le k \le 12$ are shown in Table~\ref{tabtrunc4},
and for the $L=5$ basis with $5 \le k \le 9$ in Table~\ref{tabtrunc5}. 

It is clear from Tables~\ref{tabtrunc3}--\ref{tabtrunc5} that when $L$ increases, we need higher $k$
to come close to the true, untruncated result. In other words, these data only become useful if we
are capable of extrapolating them to $k=N_L$. We have found that an efficient means of doing so
is to fit $p_N(k)$ to a second-order polynomial in the variable $x \equiv 1-\rho(k)$. Excluding gradually
data points for small $k$ and iterating the fits, as described in section~\ref{sec:extrapolslope}, we
obtain the extrapolations given as the last two entries in Table~\ref{tabpN}.
The error bar on the $L=4$ result is so small that for the purpose of the next sub-section it can be
considered essentially exact, whereas the $L=5$ result is not sufficiently precise to be used in
the subsequent analysis.

\begin{table}
\begin{center}
\caption{Determination of $p_N$ for the straight $L=4$ basis by truncation.}
\label{tabtrunc4}
\begin{tabular}{rll}
$k$ & $p_N(k)$ & $\rho(k)$ \\
\hline
  5 & 0.131331472620 & 0.871031 \\
  6 & 0.116222782813 & 0.946644 \\
  7 & 0.111355042618 & 0.981002 \\
  8 & 0.109768185411 & 0.994137 \\
  9 & 0.109296182921 & 0.998422 \\
10 & 0.109165302878 & 0.999627 \\
11 & 0.109134030810 & 0.999922 \\
12 & 0.109127261907 & 0.999985 \\
\end{tabular}
\end{center}
\end{table}

\begin{table}
\begin{center}
\caption{Determination of $p_N$ for the straight $L=5$ basis by truncation.
}
\label{tabtrunc5}
\begin{tabular}{rll}
$k$ & $p_N(k)$ & $\rho(k)$ \\
\hline
  5 & 0.198903641300 & 0.533089 \\
  6 & 0.165041787651 & 0.699171 \\
  7 & 0.135107684705 & 0.826880 \\
  8 & 0.11967129         & 0.910856 \\
  9 & 0.11                     & 0.958797 \\
  \end{tabular}
\end{center}
\end{table}

\subsubsection{Final value of $p_N$.}

To obtain a final value of $p_N$ from the finite-size data $p_N(L)$ in Table~\ref{tabpN}, we 
first observe that the subsequence with even (resp.\ odd) $L$ appears to be monotonically
increasing (resp.\ decreasing). Assuming this to be true in general, we obtain the bound
\begin{equation}
 0.109126 \le p_N \le 0.109382
\end{equation}
determining $p_N$ to within $2.5 \times 10^{-4}$. This is already competitive with the best
available numerical result \cite{Hasenbusch2008}, in which $p_N$ is estimated as $0.10919(7)$,
that is, to within a range of $1.4 \times 10^{-4}$ (up to the confidence level applied in that study).

\begin{figure}[tb]
\begin{center}
\includegraphics[width=80mm]{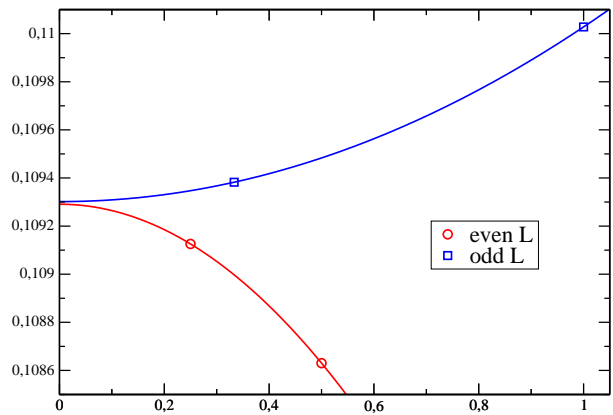}
\end{center}
\caption{The finite-size estimates $p_N(L)$ of Table~\ref{tabpN} plotted against $1/L$.
The subsequences with even and odd $L$ are extrapolated separately to the $L \to \infty$
limit, as explained in the main text.}
\label{fig:finalpc}
\end{figure}

A more precise determination can be obtained as follows. Recall that when studying
the slope at the pure Ising point, the finite-size correction was found to be of the form $a L^{-\omega}$
with $\omega = 2$; see eq.~(\ref{conjFSSexp}). Due to the irrelevance of the disorder around the
pure Ising point, we expect that $\omega = 2$ can also be applied to the data in Table~\ref{tabpN}.
Fitting the even points $L=2,4$ and the odd points $L=1,3$ separately to the form
\begin{equation}
 p_N(L) = p_N + a L^{-2}
 \label{extrapolFSSpc}
\end{equation}
we obtain $p_N^{\rm even} = 0.109291$ from the even data and $p_N^{\rm odd} = 0.109301$ from the odd date.
This is shown in Figure~\ref{fig:finalpc}. The excellent agreement between these two values
is a strong validation of the assumption (\ref{extrapolFSSpc}). For our final estimate of $p_N$
we take $p_N^{\rm even}$ as the central value (since it corresponds to the largest values of $L$),
and we conservatively estimate the error bar as twice the difference $|p_N^{\rm even} - p_N^{\rm odd}|$.
We thus arrive at the value given in the abstract:
\begin{equation}
 p_N = 0.10929 (2) \,.
 \label{pNIsingfinal}
\end{equation}
This is three or four times more accurate than the result of \cite{Hasenbusch2008}.

\subsection{Ground state in spin glasses}

The criterion (\ref{Aeq2}) leads to very accurate estimations of the phase boundary in the high-temperature
region (above the Nishimori line), as witnessed by Figure~\ref{phase-diagram} and its left inset in particular.
However, the fact that all curves end at the origin does not, at first sight, seem consistent with the
expected zero-temperature critical point at $p_0$ (see Figure~\ref{paper_fig1}). To lift this apparent
discrepancy we now study analytically the slope of the curves (\ref{Aeq2}) at the origin, using bases of arbitrarily
large size.

Let us consider the asymptotic structure of the phase diagram by applying our method to the ground
state ensemble. Eq.~(\ref{Aeq2}) can be represented by the ratios of the partition functions with
different periodic conditions as 
\begin{equation}
\left[ \log \left( 1 + \frac{Z_{+-}}{Z_{++}}+ \frac{Z_{-+}}{Z_{++}}+ \frac{Z_{--}}{Z_{++}}\right)\right] = \log 2 \,.
\end{equation}
Once we accept the trivial solution $p=0$ estimated from (\ref{Aeq2}) for bases of small size,
let us expand the above equality in $p \ll 1$ as
\begin{eqnarray}\nonumber
& & \log \left( 1 + \frac{Z^{\rm F}_{+-}}{Z^{\rm F}_{++}}+ \frac{Z^{\rm F}_{-+}}{Z^{\rm F}_{++}}+ \frac{Z^{\rm F}_{--}}{Z^{\rm F}_{++}}\right) \\
& & \quad + \sum p \log \left( 1 + \frac{Z^{{\rm AF}_1}_{+-}}{Z^{{\rm AF}_1}_{++}}+ \frac{Z^{{\rm AF}_1}_{-+}}{Z^{{\rm AF}_1}_{++}}+ \frac{Z^{{\rm AF}_1}_{--}}{Z^{{\rm AF}_1}_{++}}\right) + O(p^2) = \log 2 \,, 
\label{GS}
\end{eqnarray}
where we have omitted the factor $(1-p)^{N_L-1}$ which is not relevant for the following discussion.
Just like in the calculations in section~\ref{sec:extrapolslope}
of the slope value near the pure Ising point, the summation in (\ref{GS}) is over all configurations
with only a single antiferromagnetic interaction.

In the ground state ($K \to \infty$), the partition function on the basis reduces to the exponential
of the ground state energy times the inverse temperature. An antiperiodic boundary condition
in the horizontal (resp.\ vertical) direction can equivalently be produced by introducing a column
(resp.\ row) of antiferromagnetic horizontal (resp.\ vertical) bonds and switching back to periodic
boundary conditions. Thus, the ratios of partition functions for a basis of linear size $L$ read
\begin{eqnarray}
 Z^{\rm F}_{+-}/Z^{\rm F}_{++} &=& Z^{\rm F}_{-+}/Z^{\rm F}_{++} = \exp(-2KL) \,, \nonumber \\
 Z^{\rm F}_{--}/Z^{\rm F}_{++} &=& \exp(-4KL) \,.
\end{eqnarray}
Hence the first term in eq.~(\ref{GS}), of order $p^0$, decays exponentially when $L \to \infty$.

The second term in eq.~(\ref{GS}) can be split into two parts, depending on whether the
frustrated bonds of the disorder configuration do or do not overlap with
the horizontal row / vertical column of antiferromagnetic interactions.
In the former case we have $Z^{{\rm AF}_1}_{++} = \exp(4L^2 K -2K)$,
while $Z^{{\rm AF}_1}_{+-} = \exp(4L^2 K - 2K(L-1))$ and
$Z^{{\rm AF}_1}_{-+} = \exp(4L^2 K - 2KL -2K)$ or vice versa,
and $Z^{{\rm AF}_1}_{--}=\exp(4L^2 K-2K(L-1) -2KL )$, where $N_L = 2L^2$ is the
number of all bonds in the basis and the $4L^2$ denotes the trivial ground state energy
corresponding to ferromagnetic interactions only.
\begin{figure}[tb]
\begin{center}
\includegraphics[width=70mm]{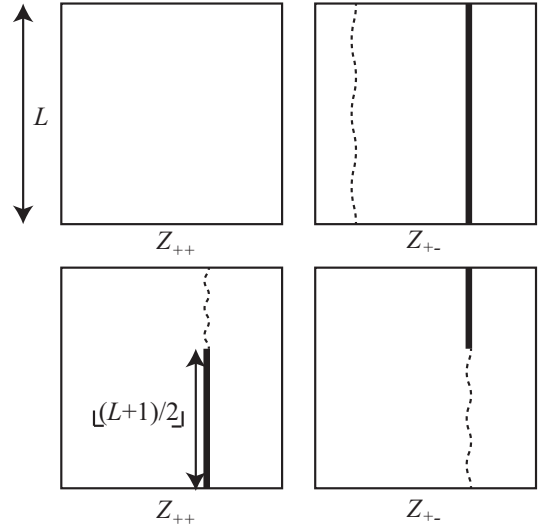}
\end{center}
\caption{Possible excitations by the induced antiferromagnetic interactions and the vertical/horizontal sequence of the antiferromagnetic interactions.
The bold lines denote the antiferromagnetic interactions, while the waving lines represent the excited bonds.
The upper panels show the typical instance of the excitations induced disorder and the sequence of the antiferromagnetic interactions.
The lower panels describe an extreme case that the induced disorder completely overlaps with the sequence of the antiferromagnetic interactions and the excitation energy under the periodic boundary condition exceeds that under the anti-periodic condition in one direction.
}
\label{GSfig}
\end{figure}
In the latter case, the partition functions meanwhile read
\begin{eqnarray}
 Z^{{\rm AF}_1}_{++} &=& \exp(4L^2 K -2K) \,, \nonumber \\
 Z^{{\rm AF}_1}_{+-} &=& Z^{{\rm AF}_1}_{-+}= \exp(4L^2 K - 2KL -2K) \,, \\
 Z^{{\rm AF}_1}_{--} &=& \exp(4L^2 K-4KL -2K) \,. \nonumber
\end{eqnarray}
Thus the second term in eq.~(\ref{GS}), of order $p^1$, also decays exponentially.

Let us repeat the same assessment for the higher-order terms, of order $p^k$, arising in
the expansion of eq.~(\ref{Aeq2}).
The most important quantity is the difference between the ground state energy with different boundary conditions.
For large enough $L$, each term vanishes exponentially, since the ground state energy with periodic
boundary conditions in both directions is lower than those with the other types of boundary conditions.
In other words, the rigidity of the ground state holds even upon introduction of a few antiferromagnetic interactions.

The first relevant term appears at order $k = \lfloor (L+1)/2 \rfloor$, where $\lfloor x \rfloor$ denotes the
integer part of $x$, since the ground state energy with periodic boundary conditions in both directions
begins to exceed the energy with anti-periodic conditions in one direction.
This is illustrated in Figure \ref{GSfig}. Therefore 
\begin{eqnarray}
& & \left(
\begin{array}{c}
L \\
\lfloor (L+1)/2 \rfloor
\end{array}
\right)p^{\lfloor (L+1)/2 \rfloor} K + O(p^{\lfloor (L+1)/2 \rfloor+1})= \log 2 \,,
\end{eqnarray}
which reads $T_c \sim c p^{L/2}$, where the coefficient $c$ is non-vanishing in the limit of large $L$.
This means that the phase boundary at the origin becomes arbitrarily flat when $L \to \infty$.
This is compatible with a scenario in which the phase boundary degenerates into a horizontal line
segment extending from the origin to the point $p_0 > 0$, as in Figure~\ref{paper_fig1}.
In other words, the duality analysis with real-space renormalization on the infinite-size basis
is consistent with the well-known structure of the phase boundary in the ground state.

Let us consider the phase boundary in the ground state in more detail.
Increasing the number of antiferromagnetic interactions yields non-trivial ground states,
which results in the disordered state. The difference of ground state energy between
the different boundary conditions then becomes $O(1)$ instead of $O(L)$.
The terms contributing to eq.~(\ref{Aeq2}) in the ground state come from such non-trivial configurations.
We can therefore write
\begin{equation}
\sum_{L>\lfloor (L+1)/2 \rfloor} S_L p^L \approx \log 2 \,,
\end{equation}
where $S_L$ denotes the number of non-trivial configurations of the antiferromagnetic interactions,
and we have omitted the insignificant coefficient coming from the difference of the ground state energy
between different boundary conditions. We expect $S_L \sim \exp(\gamma_L(p) L)$ for some finite
$\gamma_L$. In the thermodynamical limit, $L \to \infty$, the sum can be replaced by an integral:
\begin{equation}
\int_{L/2}^\infty {\rm d}L \, \exp\left\{ L\left(\gamma_L(p) + \log p\right)\right\} \approx \log 2 \,.
\label{saddlepointint}
\end{equation}

We may employ the saddle-point method to obtain the non-trivial solution $p_0$.
The saddle point $\alpha_L(p_0)$ is found by deriving the exponent of
the above integral, $\partial \gamma_L / \partial p - 1/p = 0$. This implies
\begin{equation}
\gamma_L(p_0) + \log p_0 = 0 \,,
\end{equation}
since the right-hand side of (\ref{saddlepointint}) is $O(1)$.

This means that the non-trivial solution $p_0$ is determined by the maximum of the number of non-trivial configurations of antiferromagnetic interactions in the given basis with size $L$.
The recent study by Miyazaki \cite{Miyazaki2013} reveals that the location of the critical point in the ground state is closely related to the typical value (expectation) of the frustration in the two-dimensional system, given the fraction of antiferromagnetic interactions $p$.
This author therefore considered the derivative of the expectation of the frustration with respect to the fraction $p$.
The non-trivial point in the ground state is located at a special point in which the derivative takes a special value.

Notice that the consequence of the duality analysis with real-space renormalization (\ref{Aeq2}) can be rewritten in terms of the frustration entropy by use of the gauge transformation \cite{Nishimori2001,Ohzeki2009a,Nishimori1986} as 
\begin{equation}
\frac{1}{2^N(2\cosh K(p))^{N_B}} \sum_{f} \left(\tilde{Z}^{({\rm PB})} \log Z^{({\rm PB})} - \sum_{\tau_x,\tau_y} \tilde
{Z}^{({\rm PB})}_{\tau_x,\tau_y} \log Z^{({\rm PB})}_{\tau_x,\tau_y} \right) = \log 2 \,,
\end{equation}
where $\tilde{Z}^{({\rm PB})}$ denotes the partition function with $K(p)$. 
The summation is taken over all frustration configurations. 
Further study in this direction may merge with the recent analysis of the frustration.

\subsection{Phase diagram of Potts gauge glasses}

We can extend the main parts of our analysis to the $q$-state Potts gauge glass. For simplicity
we focus on the values $q=3$ and $q=4$.

\begin{figure}[tb]
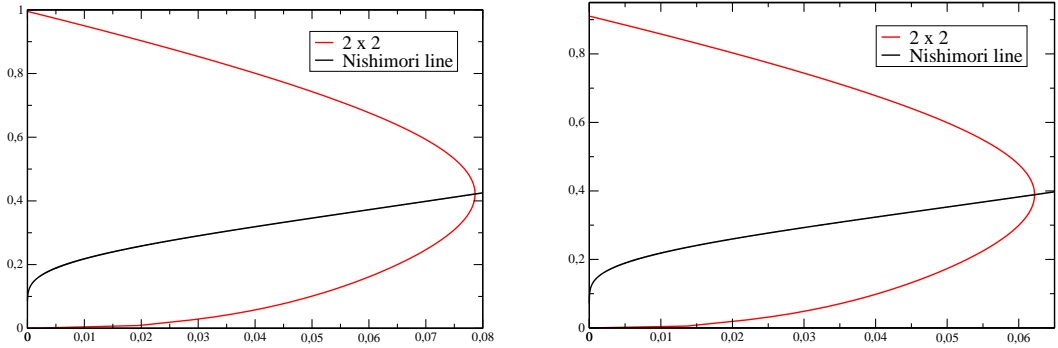

\begin{center}
\includegraphics[width=65mm]{Potts_Q3.eps} \quad \quad
\includegraphics[width=65mm]{Potts_Q4.eps}
\end{center}
\caption{Phase diagrams of the $q$-state Potts gauge glass with $q=3$ (left panel) and $q=4$ (right panel),
using the straight $2 \times 2$ basis. The Nishimori lines (\ref{NlinePotts}) are shown as solid black curves.}
\label{fig:pdq34}
\end{figure}

The phase diagrams using the $L=2$ straight basis with $N_L = 2 L^2$ bonds are shown in
Figure~\ref{fig:pdq34}.

\begin{table}
\begin{center}
\caption{Estimations of the slope $\alpha$ from eq.~(\ref{DTc}) for the Potts gauge glass with $q=3$ and $4$.}
\label{table3}
\begin{tabular}{cll}
$L$ & $\alpha^{(q=3)}$ & $\alpha^{(q=4)}$ \\
\hline
1  & $4.611745072516 $ & $5.838306370025 $\\
2  & $4.433594631195 $ & $5.574506278394$\\
3  & $4.401119996165$ & $5.522053809017$\\
4  & $4.389065445497$ & $5.501231271824$\\
5  & $4.383148271677 $ & 
\\
\end{tabular}
\end{center}
\end{table}

Estimations of the slope at the pure Potts point, $\alpha = \frac{1}{T_c} \frac{{\rm d}T}{{\rm d}p}$,
have been obtained from diagonal bases of size $L$, and are displayed in Table~\ref{table3}.
We here again assume that the slope takes the following asymptotic form 
\begin{equation}
\alpha(L) = \alpha + aL^{-\omega} \,.
\end{equation}
Extrapolations, using the same procedure as outline above, lead to the following
values for the $3$-state Potts gauge glass:
\begin{eqnarray}
\alpha^{(q=3)} = 4.374(1) \,,  \qquad a^{(q=3)} = 0.238(1) \,, \qquad \omega^{(q=3)} = 1.99(2) \,.
\end{eqnarray}
We again find an exponent $\omega^{(\rm PB)} \simeq 2$, in agreement with the conjecture
(\ref{conjFSSexp}). This provides strong evidence that this exponent is independent of $q$.

On the other hand, fitting the results for the $4$-state Potts gauge glass leads to
\begin{eqnarray}
\alpha^{(q=4)} = 5.473(2) \,, \qquad a^{(q=4)} = 0.364(2) \,, \qquad \omega^{(q=4)} = 1.84(3) \,.
\end{eqnarray}
Taken at face value, the value of $\omega^{(q=4)}$ does not support the conjecture $\omega^{(\rm PB)}=2$.
This discrepancy is however likely to be due to the strong logarithmic corrections present in the
pure $q=4$ Potts model. Note also that Table~\ref{table3} contains less data for $q=4$.

The slope $\alpha^{(q=3)}$ can be compared with the results for the phase boundary given
in Table 1 of Ref.~\cite{Jacobsen2002}. Assuming a linear phase boundary
between $p=0$ and $p=0.01$ one obtains the value $\alpha^{(q=3)} = 4.49 (5)$, where we
have multiplied the statistical error by a factor of ten to take into account that the linearity
assumption is a rather crude approximation. Our extrapolated value is much more precise than
this numerical result.

Determinations of the Nishimori points are given in Table~\ref{tab:pN34}. We now move back
to straight bases with $N_L = 2 L^2$ bonds.
For the $q=3$ case the $L=3$ entry was obtained by extrapolation of the truncated
results reported in Table~\ref{tabtrunc3q3}, by using the methodology established in
section~\ref{subsec:trunc}.%
\footnote{Note in particular that one must replace $p$ by $(q-1)p$ in eq.~(\ref{fracstatweight}) to
obtain $\rho(k)$.}
We did not investigate the size $L=3$ for $q=4$, or higher $L$ for $q=3$,
since in both cases the summation over $q^{N_L}$ disorder realizations would have
required more substantial numerical resources.

\begin{table}
\begin{center}
\caption{Determination of the Nishimori point for the $q=3$ and $q=4$ Potts gauge glasses,
using straight bases of size $L$.}
\label{tab:pN34}
\begin{tabular}{lll}
$L$ & $p_N ^{(q=3)}$ & $p_N^{(q=4)}$ \\
\hline
 1 & $0.079730752$ & $0.063096541$ \\
 2 & $0.078642990$ & $0.062210717$ \\
 3 & $0.07914779 (2)$ & \\
\end{tabular}
\end{center}
\end{table}

\begin{table}
\begin{center}
\caption{Determination of $p_N$ for the $q=3$ Potts gauge glass, 
using truncation with the straight $L=3$ basis.
The statistical sum is truncated to at most $k$ antiferromagnetic couplings.
The corresponding fraction of the total statistical weight at $p = 0.079$ is denoted $\rho(k)$.}
\label{tabtrunc3q3}
\begin{tabular}{rll}
$k$ & $p_N(k)$ & $\rho(k)$ \\
\hline
  5 & 0.084991110 & 0.947726347 \\
  6 & 0.080718923 & 0.984400423 \\
  7 & 0.079520998 & 0.996197852 \\
  8 & 0.079222147 & 0.999241785 \\
  9 & 0.079159923 & 0.999876440 \\
10 & 0.079149405 & 0.999983623 \\
\end{tabular}
\end{center}
\end{table}

The (admittedly very limited) data in Table~\ref{tab:pN34} is compatible with a scenario where
the finite-size dependence of $p_N(L)$ would be monotonically decreasing for odd $L$ and
monotonically increasing for even $L$, just like we have seen for the $q=2$ Ising case.
If we suppose this to be so, the $L \to \infty$ limit of $p_N(L)$ should satisfy the bounds
\begin{equation}
 0.07864 \le p_N^{(q=3)} \le 0.07915 \,.
\end{equation}
This is consistent with the estimate $p_N^{(q=3)} = 0.0785 (10)$ coming from numerical computations
\cite{Jacobsen2002}.

We have seen above that the finite-size dependence of the slope at the pure Potts point is
compatible with an exponent $\omega = 2$. Boldly fitting the data with $L=1$ and $L=3$ to
the form (\ref{extrapolFSSpc}) leads to
\begin{equation}
 p_N^{(q=3)} = 0.07907 \,,
\end{equation}
where we have refrained from giving an error bar, but it seems reasonable to assume it to be of the
same order of magnitude as that appearing in (\ref{pNIsingfinal}). We do however wish to stress
that for $q>2$ the disorder is relevant, and the assumption that the finite-size scaling exponent
$\omega$ might be the same as at the pure Potts point would need additional justification.

\section{Conclusion}

In this paper we have proposed a new framework for duality analysis with real-space renormalization,
by combining the existing approach \cite{Ohzeki2009a} with the graph polynomial method
\cite{Jacobsen2012,Jacobsen2013,Jacobsen2014}. We have applied our method to a variety
of two-dimensional spin glass models, namely the $\pm J$ random-bond Ising model, the
bond-diluted Ising model, and the $q$-state Potts gauge glass.

In all cases we found agreement with the existing knowledge about the phase boundary, and
we were able to substantially improve on the precision of the determination of its salient features,
such as the the location $p_N$ of the Nishimori point and the slope $\alpha$ at the pure critical
point. The shortcomings of the previous method \cite{Ohzeki2009a}, namely its slow convergence
towards $\alpha$ and its failure to predict the zero-temperature critical point $p_0$, were
dispelled, in the latter case using an analytical argument.

We should probably insist that all the results reported in this paper were obtained by rather
modest numerical means, that is, in particular, without resorting to parallelized computations. If more substantial
means were applied---say, comparable to those usually employed in numerical studies of
spin glasses---we believe that it would be fully realistic to determine $p_N$
for the $\pm J$ Ising model to a precision of $10^{-6}$.

The analysis of finite-dimensional spin glasses has been steadily and gradually developing
over recent years. We believe that the present method opens some new perspectives.
It also inevitably raises a number of open questions, to be addressed in future research.

For instance, our approach was successful in determining very accurately the phase diagram of the
$q$-state Potts gauge glass (PGG). On the other hand, it does not apply---at least not in its present form---to
the more well-studied random-bond Potts model (RBPM) \cite{Jacobsen1997,Jacobsen1998}.
For fixed $q$, both models possess a phase diagram with two free parameters: the temperature
and the strenght of the disorder. The critical point $(p_N,T_N)$ of the PGG can be fixed by
imposing two constraints: the Nishimori gauge symmetry, and the (extended) duality symmetry.
The duality symmetry is not at all obvious to establish from the microscopic formulation of the model,
and it is the object of our method. By contrast, for the RBPM it is easy to impose duality ``on average''
by choosing a self-dual distribution of the random bonds \cite{Jacobsen1997,Jacobsen1998}, but we do not know of a second constraint
that would fully fix the critical point. In particular, for any self-dual choice of the bonds, our criterion
(\ref{generalPottsresult}) is trivially satisfied, and hence does not contribute to determining the phase
boundary.

Another interesting two-dimensional spin glass model is that of the bond-disordered O($n$) model
\cite{Shimada2014}. For fixed $n$ it again has two free parameters. But in this case, no constraint
is known that would narrow in the non-trivial critical point, and accordingly its numerical study
\cite{Shimada2014} is considerably harder than that of the RBPM. Unfortunately, both the present method
and the (generalized) duality methods that it derived from \cite{Jacobsen2012,Jacobsen2013,Jacobsen2014}
are presently limited to Potts model. For the O($n$) model we do not know of any duality symmetry,
not even in the pure case.

Let us finally remark that it seems exceedingly hard to provide any operational formulation of
the $q$-state Potts gauge glass for continuous values of $q$ \cite{Jacobsen2002}. Therefore,
the interesting region of $q > 4$ \cite{Jacobsen1997,Jacobsen1998} is hard to attain, because
of the large number $q^{N_L}$ of disorder configurations that must be summed over. In particular,
the interesting limit $q \to \infty$ \cite{Picco2000,Igloi2001} is outside the scope of the current method.

\section*{Acknowledgments}

The authors are grateful for the fruitful discussions with Koji Hukushima, Ryoji Miyazaki,
Christian R.~Scullard, and Koujin Takeda.
MO acknowledges financial support from the JSPS Core-to-Core program
{\em Non-equilibrium dynamics of soft matter and information} and from
MEXT in Japan, Grant-in-Aid for Young Scientists (B) No.~24740263.
He also thanks the \'Ecole Normale Sup\'erieure and the University of Tokyo
for hospitality during this work.
The research of JLJ was supported by the Agence Nationale de la Recherche
(grant ANR-10-BLAN-0414: DIME) and the Institut Universitaire de France.

\appendix

\section{Generalizing the computation of $x_0^{({\rm PB})}$ and $x_0^{*({\rm PB})}$}
\label{appA}

The aim of this appendix is to derive the criterion (\ref{generalPottsresult}).
To this end we consider the generalization of the computation of
the principal Boltzmann factors $x_0^{({\rm PB})}$ and $x_0^{*({\rm PB})}$
by use of a basis of arbitrary size $L$.

\begin{figure}[tb]
\begin{center}
\includegraphics[width=0.9\textwidth]{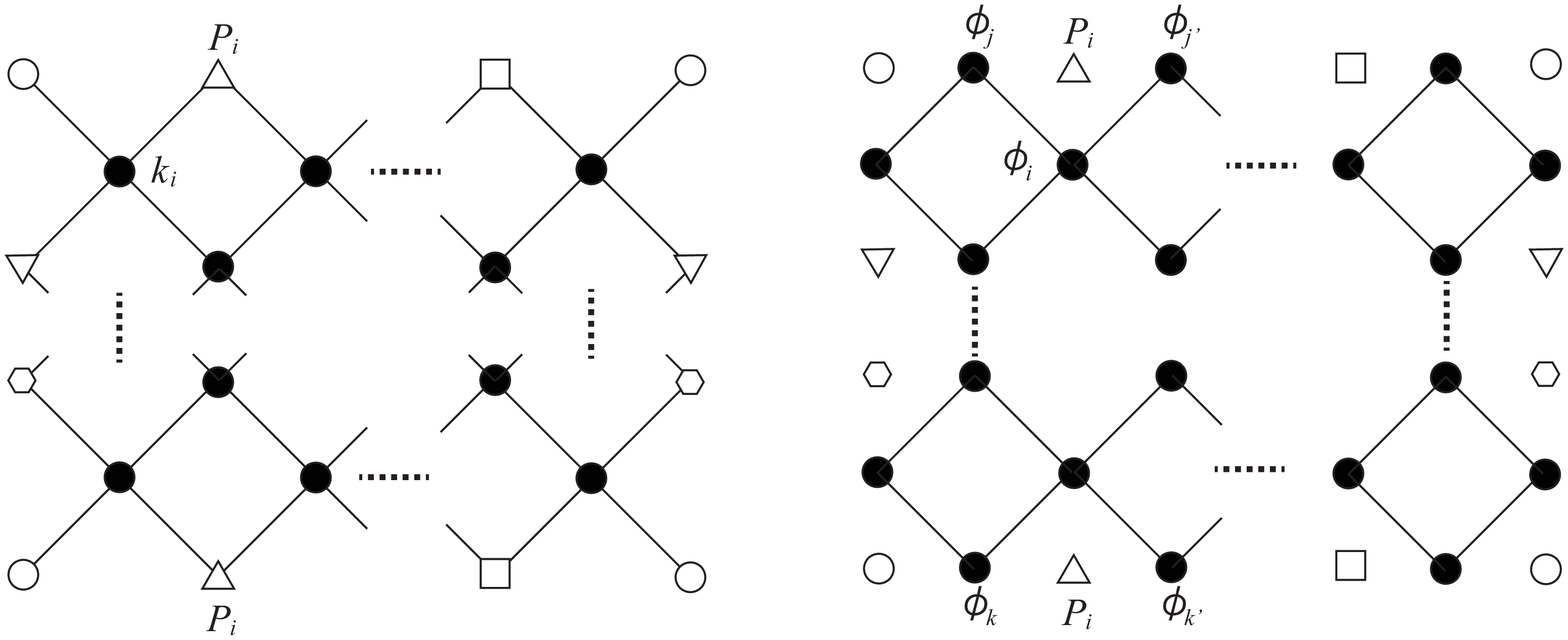}
\end{center}
\caption{ General basis with arbitrary size of $L$.
The left panel shows the basis used for evaluating $x^{({\rm PB})}_0$ and $x_0^{*({\rm PB})}$.
Here PB stands for periodic boundary conditions, meaning that matching symbols in the figure are
identified.
The boundary spin $P_b$ is set to a fixed value (zero) in the original duality with real-space renormalization.
We remove this restriction when evaluating $x^{({\rm PB})}_0$ and $x_0^{*({\rm PB})}$.
The right panel illustrates the dual basis, and the corresponding basis is that of $Z^{({\rm PB})}_{\tau_x,\tau_y}$.}
\label{AP1}
\end{figure}

The following considerations apply to any $q$-state spin model, including the
$q$-state Potts and $q$-state clock models.
The basis is as shown in Figure \ref{AP1}.
If we change the boundary condition to the periodic in both of the directions, we straightforwardly obtain
\begin{equation}
x^{({\rm PB})}_0 = Z^{({\rm PB})}_{00} \,,
\end{equation}
where we recall that $Z^{({\rm PB})}_{\tau_x,\tau_y}$ denotes the partition function with twisted periodic
boundary conditions $(\tau_x,\tau_y)$.

On the other hand, the evaluation of the dual principal Boltzmann factor under the periodic boundary conditions is not so simple.
As a warmup, we first show that the dual principle Boltzmann factor under {\em fixed} boundary condition coincides with the partition function on the graph dual to the basis.
The dual principal Boltzmann factor is defined as
\begin{eqnarray}
x^{*({\rm F})}_0 &=& \left[\left( \left(\frac{1}{\sqrt{q}}\right)^{4L^2}\sum_{\{k_i\}} \prod_{\langle ij \rangle}x^*_{k_i - k_j} \right)^n\right] \,,
\end{eqnarray}
where the superscript $(\rm F)$ denotes the fixed boundary condition.
The dual spin variable is represented by $k_i = 0,1,\ldots, q-1$.
The dual edge Boltzmann factor $x_k^*$ is given by the discrete Fourier transformation as
\begin{equation}
x_k^* = \frac{1}{\sqrt{q}}\sum_{\phi}x_{\phi}\exp\left( {\rm i}\pi \phi k \right) \,.
\end{equation}
We thus rewrite the dual principal Boltzmann factor on the basis as
\begin{eqnarray}\nonumber
x^{*({\rm F})}_0 &=& \left[\left(\left(\frac{1}{\sqrt{q}}\right)^{4L^2}\sum_{\{k_i\}} \sum_{\{\phi_{ij}\}} \prod_{\langle ij \rangle} x_{\phi_{ij}} \exp\left\{ {\rm i}\pi\phi_{ij}(k_i -k_j)\right\} \right)^n\right], \\
&=& \left[\left(\left(\frac{1}{\sqrt{q}}\right)^{4L^2}\sum_{\{k_i\}} \sum_{\{\phi_{ij}\}} \prod_{ i } \prod_{ j \in \partial i }x_{\phi_{ij}} \exp\left\{ {\rm i}\pi k_i \phi_{ij} \right\} \right)^n\right] \,.
\end{eqnarray}
The symbol $j \in \partial i$ denotes the collection of spins adjacent to the spin $i$.
We perform the summation over the dual spin variable $k_i$, obtaining
\begin{eqnarray}
x^{*({\rm F})}_0
&=& \left[\left(\left(\frac{1}{\sqrt{q}}\right)^{4L^2}q^{N_L} \sum_{\{\phi_{ij}\}} \prod_{ i } \prod_{ j \in \partial i }x_{\phi_{ij}} \delta_{q}\left\{ \sum_{j \in \partial i} \phi_{ij} \right\}\right)^n\right] \,,
\end{eqnarray}
where $N_L$ is number of the internal spins, and $\delta_q(x)$ is the mod-$q$ Kronecker 
symbol defined after eq.~(\ref{H_PGG}).

To solve the Kronecker delta constraints we introduce another set of spin variables $\phi_i$ on the dual basis,
as shown in Figure~\ref{AP1}.
The fixed boundary condition makes the newly introduced spin variables free from the boundary condition.
We must take care that the degeneracy due to the introduction of the new Potts variables must be
compensated by an overall factor $1/q$.
Recalling in addition that $N_L = 2L^2 - 2L+1$, we therefore obtain 
\begin{eqnarray}\nonumber
x^{*({\rm F})}_0 &=& \left[\left( q^{-2L}\sum_{\{\phi_{i}\}} \prod_{\langle ij \rangle } x_{\phi_{i}-\phi_j}\right)^n\right] \,.
\end{eqnarray}
The right-hand side is proportional to the partition function with the original edge Boltzmann factor under
free boundary conditions. 

We next consider the dual principal Boltzmann factor under periodic boundary conditions in both of the directions.
We add the $2L-1$ boundary spins, denoted $P_i=0,1,\ldots,q-1$, so that the basis now has $N_L = 2L^2$ spins.
A calculation similar to the above leads to the following
expression of the dual principal Boltzmann factor:
\begin{eqnarray}\nonumber
x^{*({\rm PB})}_0 &=& \left[\left(\sum_{\{\phi_{ij}\}} \prod_{ i } \prod_{ j \in \partial i }x_{\phi_{ij}} \delta_{q}\left\{ \sum_{j \in \partial i} \phi_{ij} \right\} \prod_b \delta_{q}\left\{ \sum_{j \in \partial b} \phi_{bj} \right\}\right)^n\right] \,.
\end{eqnarray}
As the boundary effect, the summation over $P_i$ gives rise to another product of Kronecker deltas.
We can solve the former Kronecker delta constraints similarly by introducing further spin variables $\phi_i$ as on the dual graph.
The boundary constraints $\prod_b \delta_{q}\left\{ \sum_{j \in \partial b} \phi_{bj} \right\}$ then reads $ \phi_j - \phi_k + \phi_{j'} - \phi_k' \equiv 0$ (mod $q$).
It means that the differences between adjacent spins, $\phi_j - \phi_{j'}$ and $\phi_k - \phi_k'$ for each pair on both of the boundaries should be identical.
In other words, the constraints can be solved by twisted periodic boundary conditions for each boundary spin as $\phi_k = \phi_j + \tau_y$ where $\tau_x=0,1,\ldots,q-1$.
We thus reach
\begin{eqnarray}\nonumber
x^{*({\rm PB})}_0 &=& \left[\left(\frac{1}{q}\sum_{\tau_x,\tau_y}Z^{({\rm PB})}_{\tau_x,\tau_y}\right)^n\right] \equiv \left[\left(\frac{1}{q}Z\right)^n\right].
\end{eqnarray}
The power of $q$ has been changed from the case of the fixed boundary conditionx since the summation over the $2L-1$ boundary spin variables yields $q^{2L-1}$.

We have shown that the dual principal Boltzmann factor under periodic condition in both of the directions corresponds to the sum over partition functions with all possible boundary conditions $(\tau_x,\tau_y)$.
This is a generalization of the manipulation in Ref.~\cite{Bugrii1996} to the case of the Potts model.
This concludes the derivation of eq.~(\ref{generalPottsresult}).

\section*{References}
\bibliographystyle{iopart-num}
\bibliography{Paper_ver2}
\end{document}